\documentclass[twocolumn]{aastex631}

\usepackage{iftex}

\ifxetex
  \usepackage[silent]{xeCJK}
\else
  \usepackage{CJKutf8}
\fi

\newcommand{\CJKmulti}[1]{
  \ifxetex
    #1
  \else
  \!\begin{CJK*}{UTF8}{gbsn}#1\end{CJK*}\!\!\!
  \fi
}

\usepackage{amsmath}

\begin{document}

\title{The Dust Extinction Curve: Beyond R(V)}

\author[0000-0001-5417-2260]{Gregory M. Green}
\affiliation{Max-Planck-Institut f\"{u}r Astronomie \\
K\"{o}nigstuhl 17, D-69117 Heidelberg, Germany}

\author[0000-0003-3112-3305]{Xiangyu Zhang (\CJKmulti{张翔宇})}
\affiliation{Max-Planck-Institut f\"{u}r Astronomie \\
K\"{o}nigstuhl 17, D-69117 Heidelberg, Germany}

\author[0000-0003-1863-1268]{Ruoyi Zhang (\CJKmulti{张若羿})}
\affiliation{School of Physics and Astronomy, Beijing Normal University, \\
No.19, Xinjiekouwai St, Haidian District, Beijing 100875, People's Republic of China}
\affiliation{Max-Planck-Institut f\"{u}r Astronomie \\
K\"{o}nigstuhl 17, D-69117 Heidelberg, Germany}



\begin{abstract}

The dust extinction curve is typically parameterized by a single variable, $R(V)$, in optical and near-infrared wavelengths. $R(V)$ controls the slope of the extinction-vs.-wavelength curve, and is thought to reflect the grain-size distribution and composition of dust. Low-resolution, flux-calibrated BP/RP spectra from Gaia have allowed the determination of the extinction curve along sightlines to 130 million stars in the Milky Way and Magellanic Clouds. We show that these extinction curves contain more than a single degree of freedom -- that is, that they are not simply described by $R(V)$. We identify a number of components that are orthogonal to $R(V)$ variation, and show that these components vary across the sky in coherent patterns that resemble interstellar medium structure. These components encode variation in the 770\,nm extinction feature, intermediate-scale and very broad structure, and a newly identified feature at 850\,nm, and likely trace both dust composition and local conditions in the interstellar medium. Correlations of the 770\,nm and 850\,nm features with $R(V)$ suggest that their carriers become more abundant as the carrier of the 2175\,\AA{} feature is destroyed. Our 24~million extinction-curve decompositions and feature equivalent-width measurements are publicly available at \href{https://dx.doi.org/10.5281/zenodo.14005028}{DOI:10.5281/zenodo.14005028}.

\end{abstract}

\keywords{Interstellar dust (836) --- Interstellar dust extinction (837) --- Interstellar medium (847) --- Polycyclic aromatic hydrocarbons (1280) --- Principal component analysis (1944)}


\section{Introduction}
\label{sec:intro}

Dust extinction as a function of wavelength (the extinction curve) is diagnostic of the chemical composition and size distribution of dust grains. Historically, the shape of the extinction curve in the Milky Way has been described by a single parameter, ${R(V) \equiv A(V) / E(B\!-\!V)}$, which is the ratio of extinction in the $V$ band to reddening in ${B-V}$ bands \citep[\textit{e.g.},][]{CCM1989IROpticalUVExtinction,Fitzpatrick1999InterstellarExtinction,Gordon2023FUVtoMIRExtinctionCurve}. If this single parameter is sufficient to describe all extinction curves throughout the Galaxy, then that suggests that the composition of the dust anywhere in the Galaxy can likewise be described by a single parameter. In this picture, all dust populations would exist somewhere along a one-dimensional continuum, with one extreme being a bottom-heavy (a higher proportion of small dust grains), low-$R(V)$ population, and the other extreme being a top-heavy, high-$R(V)$ population. There is already evidence against this simplistic paradigm from the ultraviolet (UV), where $R(V)$ has been found to be insufficient to capture all Milky Way extinction-curve variation \citep{PeekSchiminovich2013UVExtinction}. Additional evidence against this paradigm comes from \citet{ZhangGreen2025ScienceRv3D}, which found that high $R(V)$ tends to occur in two very different physical regimes, the diffuse interstellar medium (ISM) and dense molecular clouds, with low $R(V)$ occurring at intermediate densities. The differing conditions in these two high-$R(V)$ regimes could leave some imprint on the extinction curve -- that is, variations that cannot be captured by $R(V)$ alone. In this paper, we use 24~million well measured optical extinction curves from \citet{ZhangGreen2025ScienceRv3D}, covering the entire low-Galactic-latitude sky, to test whether one parameter, $R(V)$, really is sufficient to describe all optical extinction curves.

Our approach is to decompose our extinction curves into a small number of components, ordered in some way by statistical significance. We expect the most significant component to represent $R(V)$ variation. Additional components may represent physical changes in dust composition, systematic errors in the data, or a combination of both.

The idea of decomposing extinction curves into separate components has been applied several times before. \citet{Massa1980UVReddeningComponents} decomposed the UV extinction curve into two components. \citet{Schlafly2016ExtinctionCurve} measured optical-NIR reddening curves (\textit{i.e.}, differences in extinction between different wavelengths) of 37,000 stars using a combination of APOGEE spectroscopy and broadband photometry from Pan-STARRS~1, 2MASS and WISE. \citet{Schlafly2017ExtinctionCurveMapping} used the second principal component from this analysis to map $R(V)$ across the APOGEE footprint. \citet[][``MFG20'']{MassaFitzpatrickGordon2020OpticalExtinctionISS} identified three statistically significant principal components in optical-UV extinction curves measured by the International Ultraviolet Explorer and Hubble/STIS. MFG20 analyzed the behavior of both the ``very broad structure'' (VBS), a depression in extinction from roughly 500--670\,nm, and ``intermediate-scale structure'' (ISS), which refers to features in the extinction curve that are wider than diffuse interstellar bands (DIBs) but narrower than the VBS. MFG20 found the strengths of the ISS features at 430, 487 and 630\,nm to be uncorrelated with $R(V)$, but found the features at 430 and 487\,nm to be correlated with the strength of the 2175~\AA{} extinction bump. More recently, \citet{MaizApellaniz2021ISMBand7700AA} identified a feature centered on 770\,nm, with a FWHM of $\sim$17.7\,nm.

Now, thanks to Gaia XP spectra \citep{Prusti2016GaiaMission,Vallenari2023GaiaDR3SummaryContent,DeAngeli2023GaiaDR3BPRPProcessingValidation,Montegriffo2023GaiaDR3BPRPExternalCalibration} we have enough measured extinction curves -- measured densely across the sky -- to not only look for statistically significant components of the extinction curve, but to also map their strengths across the sky. This allows correlation between extinction-curve variations and other physical conditions in the interstellar medium (ISM). Sky maps of extinction-curve variation also allow the separation of real, physical variations in the extinction curve from observational artifacts, based on the typical imprints (\textit{e.g}., dust-like filamentary structure vs. Gaia scanning patterns) on the sky. We believe this combination of spectral and spatial information can be a powerful tool for identifying and determining the physical origin of extinction-curve variations.

\section{Empirical Extinction Curves}
\label{sec:ext_curves}

\begin{figure*}
  \centering
  \includegraphics[width=0.95\textwidth]{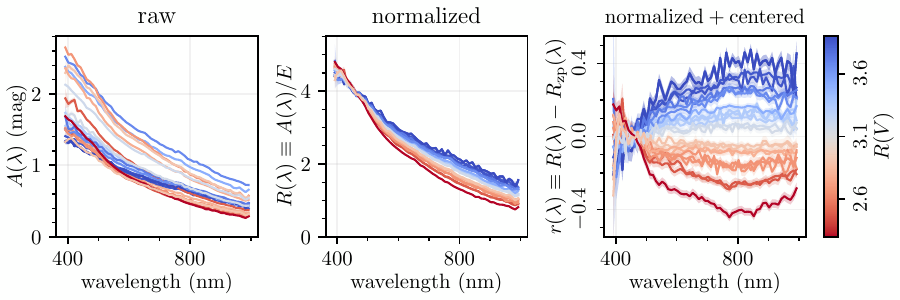}
  \caption{Normalization and centering of our empirical extinction curves. The left panel shows a small subset of our raw, empirical extinction curves, $A\left(\lambda\right)$. The middle panel shows the same extinction curves, normalized by the ZG25 estimate of scalar extinction, $E$, which is similar to $E\left(B\!-\!V\right)$. The right panel shows the normalized extinction curves, with a reference curve subtracted out. Each extinction curve is colored by its $R(V)$ estimate in ZG25, and shaded envelopes show the $\pm 1 \sigma$ uncertainties on each curve. We extract the most significant components of the right extinction curves using our CovPCA algorithm. \label{fig:A_normalization}}
\end{figure*}

\citet[][``ZG25'']{ZhangGreen2025ScienceRv3D} inferred distances and intrinsic stellar and dust parameters from Gaia XP spectra and parallaxes, augmented by near-infrared (NIR) photometry from the Two Micron All-Sky Survey \citep[][``2MASS'']{Skrutskie20062MASS} and the Wide-field Infrared Survey Explorer \citep[][``WISE'']{Wright2010WISE,Schlafly2019unWISE}, based on a data-driven forward model. This resulted in reliable measurements of dust extinction and $R(V)$ along sightlines towards 130 million stars. The ZG25 model treats the dust extinction curve as a single-parameter family of curves, which is determined directly from the data (with weak priors on the continuity of the curves in wavelength space).

However, in this paper, we intend to go beyond the assumption that the shape of the extinction curve is determined by a single parameter. We first determine an ``empirical'' extinction curve for each star by comparing the ZG25 prediction of each star's intrinsic, zero-extinction spectrum to the observed, extinguished spectrum. To do this, we input the ZG25 estimates of the stellar distance and atmospheric parameters ($T_{\mathrm{eff}}$, $\left[\mathrm{Fe/H}\right]$, $\log g$) into the ZG25 model, assuming zero extinction, obtaining a predicted intrinsic Gaia XP spectrum, $\vec{f}_{\mathrm{int}}$ (with each component representing a different wavelength), for each star. In this work, we focus on the Gaia XP spectral range of the ZG25 model, which spans 392--992\,nm, sampled every 10\,nm, for a total of 61 wavelength samples.

Next, we compare these intrinsic spectra to the observed Gaia XP spectra to obtain extinction as a function of wavelength. The observed spectral fluxes contain uncertainties, which the Gaia catalog assumes to be normally distributed. We calculate the covariance matrix of the flux uncertainties for each spectrum identically to ZG25. Including the lowest-order correction for observational uncertainty, the expectation value of extinction at a single wavelength is given by
\begin{align}
  \langle A \rangle &\simeq -\frac{2.5}{\ln 10} \left[
    \ln\left(\frac{\mu_f}{f_{\mathrm{int}}}\right)
    - \frac{1}{2}\left(\frac{\sigma_f}{\mu_f}\right)^{\! 2}
  \right]
  \, ,
  \label{eqn:A_mean}
\end{align}
where $\mu_f$ and $\sigma_f$ are the mean and standard deviation of observed flux, respectively, at the given wavelength (see Appendix~\ref{app:lnx-derivation} for a derivation). The covariance matrix of extinction at two wavelengths, $\lambda_i$ and $\lambda_j$, is approximately given by
\begin{align}
  \mathrm{cov}\!\left( A_i, A_j \right)
  \simeq \left(\frac{2.5}{\ln 10}\right)^{\!\! 2} \left(
    \frac{\sigma_{f,ij}^2}{\mu_{f,i} \mu_{f,j}}
    -\frac{1}{4}\frac{\sigma_{f,i}^2 \sigma_{f,j}^2}{\mu_{f,i}^2 \mu_{f,j}^2}
  \right) \, ,
\end{align}
where $\sigma_{f,ij}^2$ is the covariance between the observed fluxes at $\lambda_i$ and $\lambda_j$, and $\mu_{f,i}$ and $\sigma_{f,i}$ are the mean and standard deviation, respectively, of the observed flux at wavelength $i$. We thus obtain an estimate of the extinction of each star and its covariant uncertainties, which we treat as Gaussian.

As we are interested in the \textit{shape} of the extinction curve, independent of the overall amount of extinction, we normalize each extinction curve by the ZG25 estimate of the scalar amount of extinction, $E$: $R\left(\lambda\right) \equiv A\left(\lambda\right)/E$. We additionally subtract off the mean shape of the curve (averaged over the extinction curves), which we call $R_{\mathrm{zp}}\left(\lambda\right)$. We will thus look for variation in the normalized and ``centered'' extinction curves, which we denote by $r\left(\lambda\right)$:
\begin{align}
  r\left(\lambda\right)
  \equiv \frac{A\left(\lambda\right)}{E} - R_{\mathrm{zp}}\left(\lambda\right)
  \, .
\end{align}
As we work with Gaia XP spectra sampled at discrete wavelengths, we work with the vector $\vec{r}$. The covariance matrix of $\vec{r}$ is simply the covariance matrix of $\vec{A}$, scaled by $1/E^2$.

Fig.~\ref{fig:A_normalization} illustrates how we transform our raw extinction curves, $A\left(\lambda\right)$, into our normalized and centered extinction curves, $r\left(\lambda\right)$.

It is worth noting that although we are looking for higher-order variations in the extinction curve, we obtain our estimates of the intrinsic (zero-extinction) stellar spectra from a model that treats dust extinction as an $R(V)$-dependent family of curves. This intrinsic spectrum is determined by fitting the entire optical--NIR spectral range, spanning 392\,nm to $\sim 4.5\,\mathrm{\mu m}$ (WISE W2 band). In principle, variations in the extinction curve beyond $R(V)$ should affect our determination of the intrinsic stellar parameters, and therefore affect our estimate of the intrinsic spectrum and extinction curve. Our assumption is that given an accurate $R(V)$-dependent extinction curve, this 2$^{\mathrm{nd}}$-order effect should be small. This staged approach -- first using an $R(V)$-dependent extinction model to determine intrinsic spectrum, and then looking for higher-order variations in the resulting ``empirical'' extinction curves -- is similar to the approach used by \citet{Schlafly2016ExtinctionCurve}.

\section{Decomposition of Extinction Curves}
\label{sec:ext_decomposition}

\begin{figure*}
  \centering
  \includegraphics[width=0.95\textwidth]{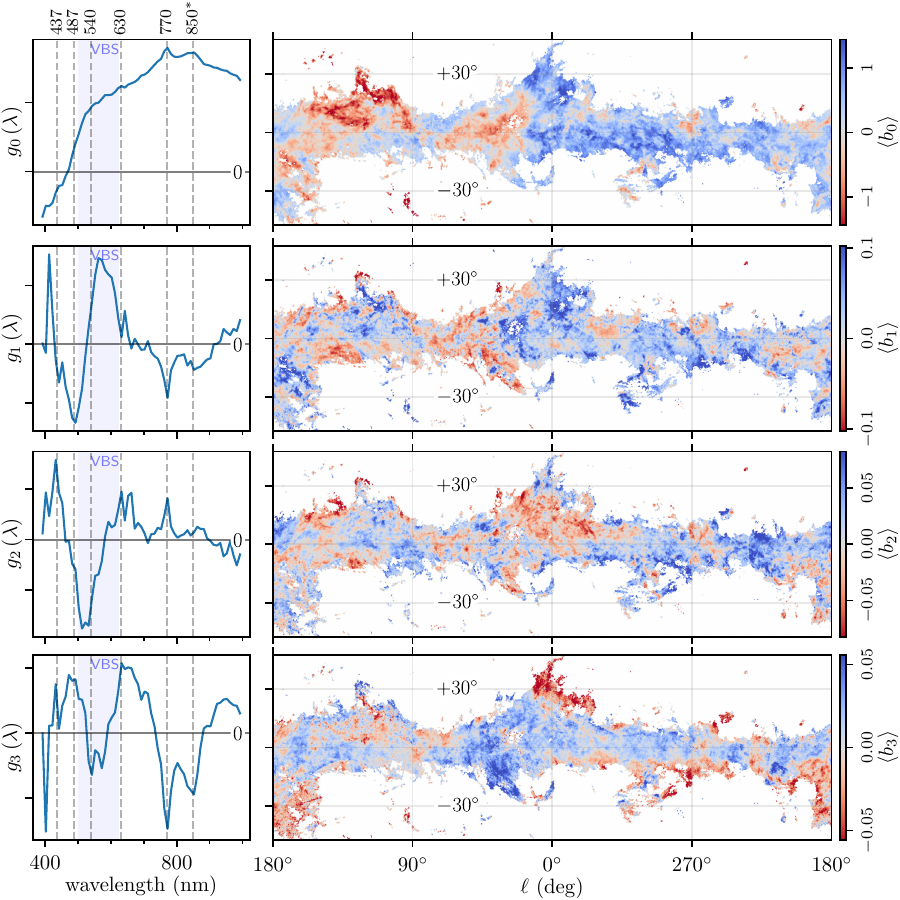}
  \caption{The first four components of the extinction curve (left panels), and corresponding sky maps of their inverse-variance-weighted mean strengths (right panels). We only show regions of the sky for which the inverse-variance-weighted mean extinction, $E$, is greater than 0.15. The sky maps of these components are relatively free of Gaia scanning-pattern artifacts, and reveal patterns that visually resemble ISM structures. Component 0 is nearly perfectly correlated with $R(V)$, while the components 1 through 3 trace orthogonal degrees of freedom in the extinction curve. In the left panels, the central wavelengths of five previously identified extinction features and one newly identified feature (at 850\,nm) are denoted by dashed lines, while the ``very broad structure'' (VBS) is indicated by a shaded blue region. These four components all modify the strength of the extinction features at 770 and 850\,nm, as well as the properties (central wavelength and depth) of the VBS. \label{fig:skymap_components_0}}
\end{figure*}

\begin{figure*}
  \centering
  \includegraphics[width=0.95\textwidth]{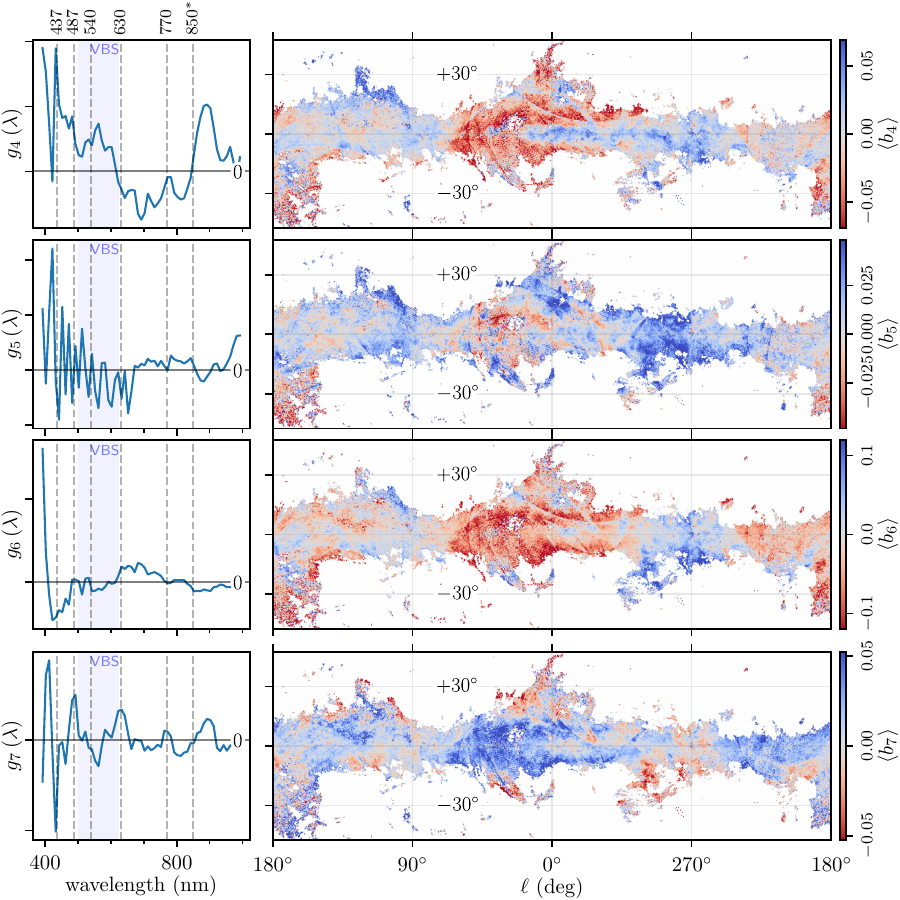}
  \caption{Components 4 through 7 of the extinction curve (left panels), along with sky maps of their mean strengths (right panels). Gaia scanning patterns are visible in these component maps, in the form of long arcs and small checkerboard patterns, though some structures that resemble ISM patterns are still visible. \label{fig:skymap_components_1}}
\end{figure*}

\begin{figure*}
  \centering
  \includegraphics[width=0.95\textwidth]{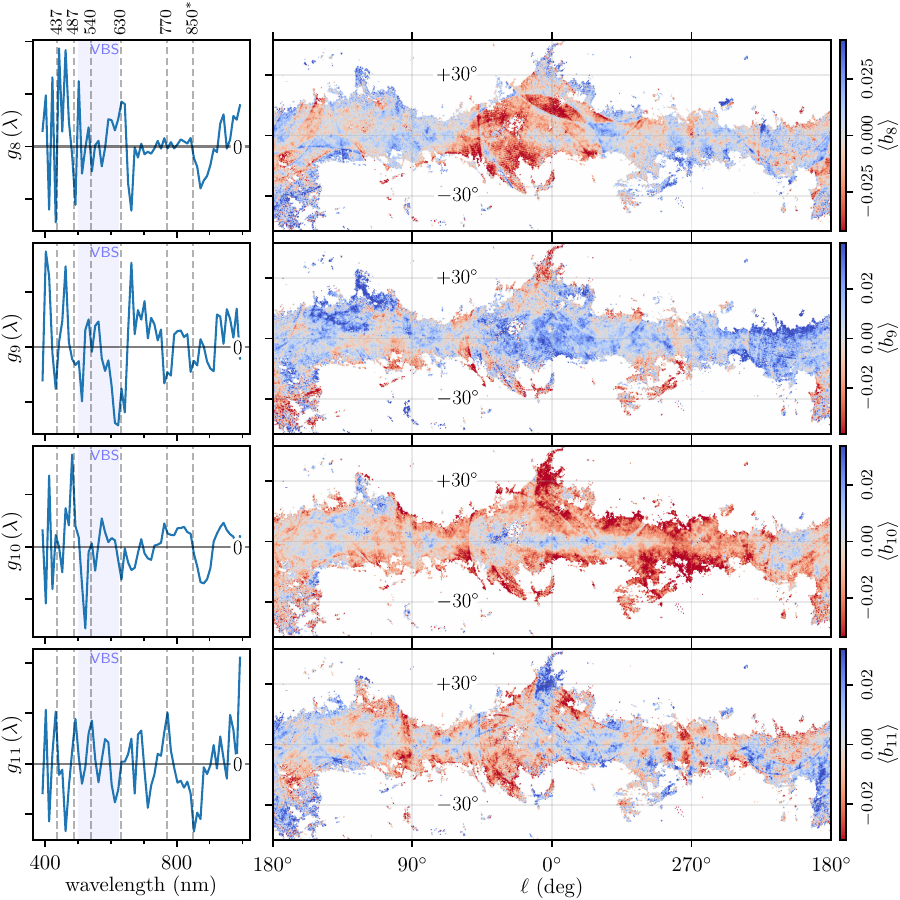}
  \caption{Components 8 through 11 of the extinction curve (left panels), along with sky maps of their mean strengths (right panels). These higher-order components show oscillatory patterns in wavelength that likely capture noise properties of Gaia, and some of the sky maps show strong Gaia scanning-pattern artifacts. Nevertheless, the sky maps of components 9 and 11 show strong ISM-like structure. Notably, component 11 may contain information about the differential strengths of the 770 and 850\,nm extinction features, which are otherwise nearly perfectly correlated. \label{fig:skymap_components_2}}
\end{figure*}

\begin{figure*}
  \centering
  \includegraphics[width=0.95\textwidth]{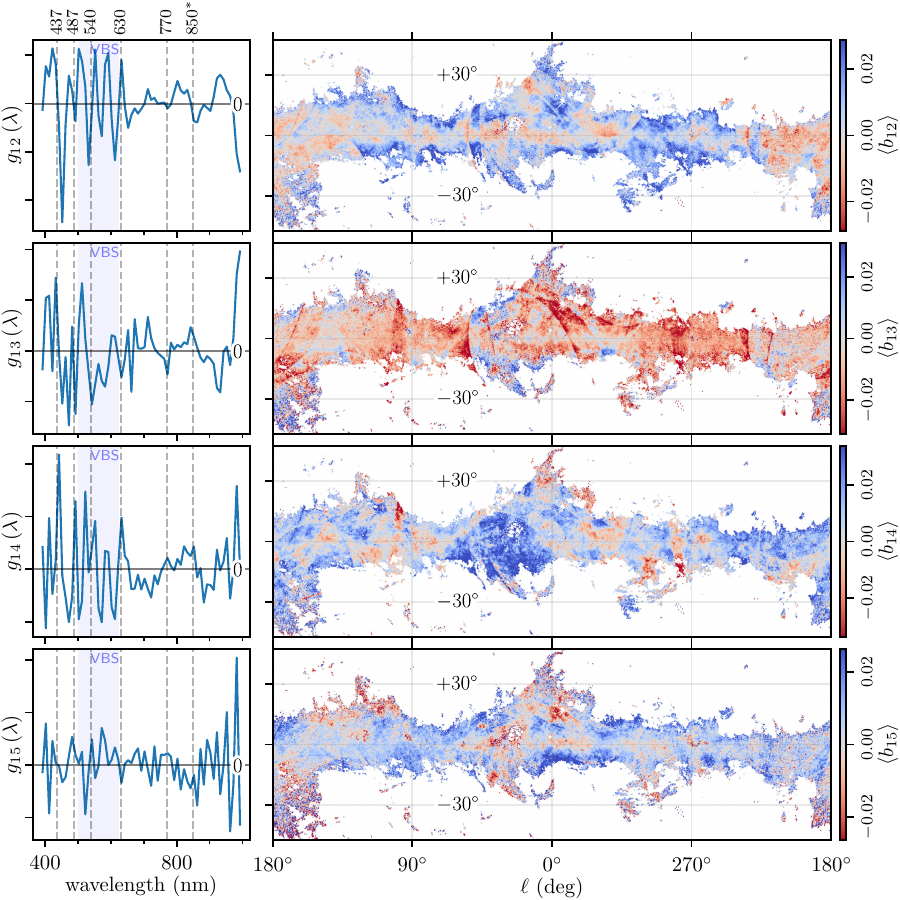}
  \caption{The last four components of the extinction curve (left panels), along with sky maps of their mean strengths (right panels). These components show oscillatory patterns in wavelength, and the sky maps show strong Gaia scanning-pattern artifacts. Some ISM-like structure is still visible in components 12, 14, and 15. \label{fig:skymap_components_3}}
\end{figure*}

\begin{figure*}
  \centering
  \includegraphics[width=0.95\textwidth]{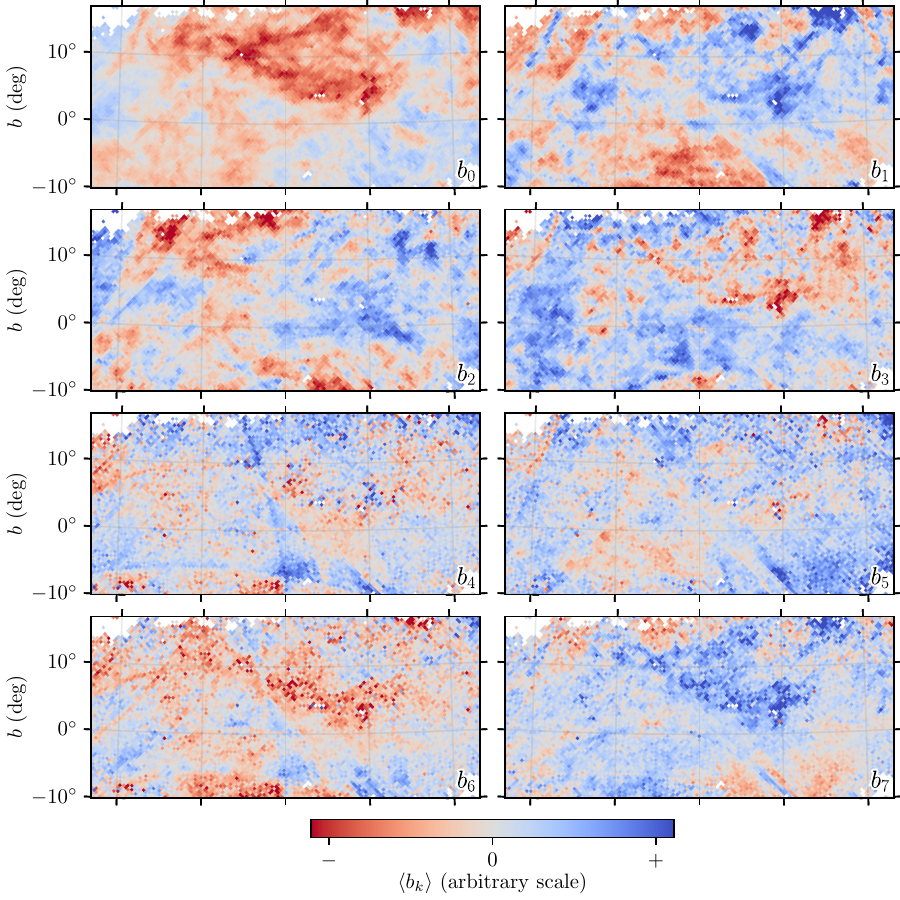}
  \caption{Maps of the mean strength of the first 8 components in the vicinity of Cepheus. The color maps are centered on zero (gray), but the scales are arbitrary, with blue indicating positive values, and red indicating negative values. Components 0-3 show spatial patterns that qualitatively resemble ISM structures, while components 4-7 show strong Gaia scanning-pattern artifacts, in the form of long arcs and checkerboard patterns. \label{fig:skymap_components_zoomin}}
\end{figure*}

Our empirical extinction curves are not guaranteed to belong to a single-parameter family, and indeed, our intent is to search for additional variation beyond $R(V)$. We search for a small number of components that, when combined linearly, can match most of our high-quality empirical extinction curves well.

We define ``high-quality'' (HQ) extinction curves based on the signal-to-noise ratio of the underlying Gaia XP spectra ($\mathtt{phot\_X\_mean\_flux}/\mathtt{phot\_X\_mean\_flux\_error} > 2500$ for \texttt{X} = \texttt{bp} and \texttt{rp}), the quality of the ZG25 stellar parameter estimates ($\mathtt{quality\_flags} < 8$ and $\mathtt{teff\_confidence} > 0.5$), and the ZG25 estimate of scalar extinction ($E > 1$). These cuts select stars for which the observed and intrinsic spectra are well determined, and for which there is enough extinction to determine the \textit{shape} of the extinction curve. In all, we obtain 29,968 HQ sources.

Our extinction curves reside in a 61-dimensional vector space (with each dimension representing extinction at a different wavelength). We look for a low-dimensional subspace that can explain most of the variation in the extinction curves. In technical terms, we search for a small ($\ll 61$) set of universal basis vectors, as well as coefficients for each extinction curve, which minimize the average $\chi^2$ difference between the extinction curves and their representation in the subspace. This is similar in spirit to Principal Component Analysis (PCA), but with the critical difference that we would like to take into account the covariant uncertainties on each observation.

We represent the basis vectors, which define the subspace, using the matrix $\mathbf{G}$, which has shape $n_{\lambda} \times n_{b}$, where $n_{\lambda} = 61$ is the number of wavelengths in each extinction curve, and $n_{b}$ is the dimensionality of our subspace (which we will set to 16 in this work). Each column of $\mathbf{G}$ is thus a basis vector. We represent the coefficients of extinction curve $i$ in this subspace by the vector $\vec{b}_i$, which is of $n_{b}$ dimensions. Extinction curve $i$ is thus to be approximated by $\vec{r}_i \simeq \mathbf{G} \, \vec{b}_i$. We seek to minimize the total $\chi^2$ distance between all the observed extinction curves and their representations in the subspace:
\begin{align}
  \chi^2 = \sum_{\mathrm{star}\ i}
    \left( \mathbf{G} \, \vec{b}_i - \vec{r}_i \right)^{\! T}
    \mathbf{C}_i^{-1}
    \left( \mathbf{G} \, \vec{b}_i - \vec{r}_i \right)
  \, ,
  \label{eqn:chi2_total}
\end{align}
were $\mathbf{C}_i$ is the covariance matrix of the uncertainties on extinction curve $i$. We iteratively solve for the basis vectors, $\mathbf{G}$, and the coefficients, $\vec{b}_i$. Holding $\mathbf{G}$ constant, there is a linear solution for each $\vec{b}_i$. Similarly, holding all of the coefficients, $\vec{b}_i$, constant, there is a linear solution for $\mathbf{G}$. Before determining the coefficients, $\vec{b}_i$, we orthogonalize the basis vectors in $\mathbf{G}$, guaranteeing an orthonormal basis set. $\chi^2$ is quadratic in the coefficients of both $\mathbf{G}$ and $\vec{b}_i$, and through repeated iterations, one approaches an optimal solution. We term this process ``CovPCA,'' and give further details in Appendix~\ref{app:cov_pca}. CovPCA is similar to the approach developed in \citet{TsalmantzaHogg2012PCASpectra}, with the difference that CovPCA takes into account covariances between measured fluxes at different wavelengths, which is critical for Gaia XP spectra.

There is an exact $O\left(n_{b}\right)$ degeneracy in the CovPCA solution: any rotation (within the subspace spanned by $\mathbf{G}$) or reflection of the coefficients $\vec{b}_i$ yields an equivalent solution (under a corresponding transformation of $\mathbf{G}$). Put more precisely, in Eq.~\eqref{eqn:chi2_total}, $\chi^2$ only depends on products of the form $\mathbf{G}\vec{b}_i$. If the matrix $\mathbf{O}$ belongs to the orthogonal group $O\left(n_{b}\right)$, then $\mathbf{O}^T = \mathbf{O}^{-1}$, so $\mathbf{G} \mathbf{O}^T \mathbf{O} \vec{b}_i = \mathbf{G}\vec{b}_i$. Transforming $\mathbf{G} \rightarrow \mathbf{G}\mathbf{O}^T$ and every $\vec{b}_i \rightarrow \mathbf{O}\vec{b}_i$ thus leaves $\chi^2$ unchanged. Additionally, unlike in PCA, the basis vectors in $\mathbf{G}$ are not ordered according to their significance.

In order to obtain a set of basis vectors ordered by significance, we take an iterative approach. We first project all of the extinction curves (and their covariance matrices) down into the 16-dimensional subspace. We then run CovPCA with one dimension to find the basis vector in the subspace that minimizes $\chi^2$. We term this most significant vector $\vec{g}_0$. We then project all of the extinction curves, as well as their covariance matrices, down to a subspace that does not span $\vec{g}_0$. We repeat this process, finding the next-most significant basis vector, $\vec{g}_1$. We iteratively repeat this procedure until we have $n_{b}$ basis vectors, which we transform back into the original space and combine into a matrix $\mathbf{G}$. This iterative process does not fundamentally alter the 16-dimensional subspace originally identified by CovPCA, but simply rotates the basis vectors so that they are ordered by significance. The left panels of Figs.~\ref{fig:skymap_components_0}--\ref{fig:skymap_components_3} show the resulting components. We then use $\mathbf{G}$ to solve for the coefficients, $\vec{b}_i$, that minimize $\chi^2$. We term this process ``iterative CovPCA.''

The left panels of Figs.~\ref{fig:skymap_components_0}--\ref{fig:skymap_components_3} plot our basis vectors, $\vec{g}_0$ through $\vec{g}_{15}$, as a function of wavelength. Note that the continuity of our basis vectors in wavelength space comes solely from the data. We do not impose any smoothness regularization on the bases. We discuss the physical meaning of these basis vectors in the following section. We find that our most significant basis vectors capture similar extinction-curve variations as the two most significant basis vectors derived by MFG20. Specifically, we find that the two most significant basis vectors in MFG20 can be closely approximated as linear combinations of our components 0 -- 3 (see Appendix~\ref{app:massa_etal_2020_comparison}).

Having determined a suitable low-dimensional basis that explains most of the variation in HQ extinction curves, we now project a larger set of medium-quality (MQ) extinction curves into this basis. We define MQ in the same way as HQ extinction curves, but with lower thresholds on signal-to-noise ($\mathtt{phot\_X\_mean\_flux}/\mathtt{phot\_X\_mean\_flux\_error} > 500$) and extinction ($E>0.1$). This yields 23.88 million MQ extinction curves. For each MQ extinction curve $i$, we calculate the coefficients, $\vec{b}_i$, that minimize $\chi^2$ (Eq.~\ref{eqn:chi2_total}), holding the basis, $\mathbf{G}$, fixed. For each extinction curve, we calculate the resulting $\chi^2$, as well as a reduced value: $\chi_{\nu}^2 \equiv \chi^2 / \left(n_{\lambda}-n_{b}\right)$. This allows us to reject extinction curves that are poorly explained by our subspace.

\section{Sky Maps of Extinction-Curve Variation}
\label{sec:sky_maps}

The right panels of Figs.~\ref{fig:skymap_components_0}--\ref{fig:skymap_components_3} show the average strength of each component across the sky. In detail, we take the inverse-variance-weighted average of each coefficient in HEALPix \citep{Gorski2005HEALPix} $\mathtt{nside}=128$ pixels ($\sim \! 27^{\prime}$ resolution). We exclude extinction curves that are poorly represented by our subspace by requiring that $\chi_{\nu}^2 < 1.5$ (which removes 1.82\% of extinction curves), and we mask regions of the sky with low average extinction ($E < 0.15$, as determined by ZG25) or fewer than 4 stars per pixel (equivalent to $\sim \! 19\ \mathrm{stars}\,\mathrm{deg}^{-2}$).

Unsurprisingly, component 0 corresponds closely to $R(V)$ as determined by ZG25, as $R(V)$ has long been known to be the dominant mode of variation in the optical extinction curve. However, the higher-order maps show variation in the extinction curve along axes that are orthogonal to $R(V)$ variation. Some of this variation -- particularly in higher-order components -- is affected by Gaia systematics, as can be seem from sharp arcs and small checkerboard patterns in some of the sky maps. However, the patterns in the four lowest-order component maps (Fig.~\ref{fig:skymap_components_0}) strongly resemble interstellar medium cloud structure in a qualitative sense, suggesting that these components trace real variation in the extinction curve. Fig.~\ref{fig:skymap_components_zoomin} shows zoom-ins of our component maps on the vicinity of Cepheus. We see fine, ISM-like angular structure in the first four components, and strong Gaia scanning-pattern artifacts mixed with some amount of ISM-like structure in the higher-order components.

\subsection{Robustness of the extinction-curve components}
\label{sec:robustness}

\begin{figure*}
  \centering
  \includegraphics[width=\textwidth]{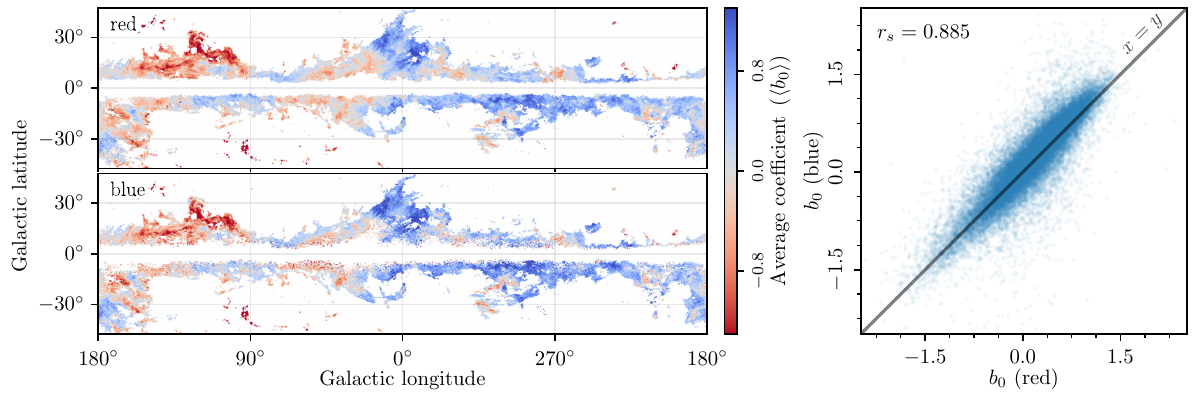}
  \includegraphics[width=\textwidth]{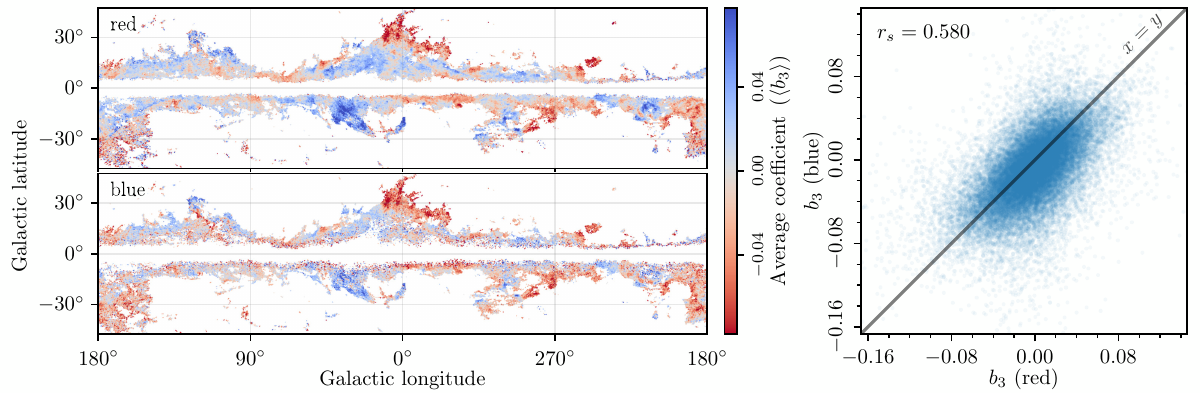}
  \includegraphics[width=\textwidth]{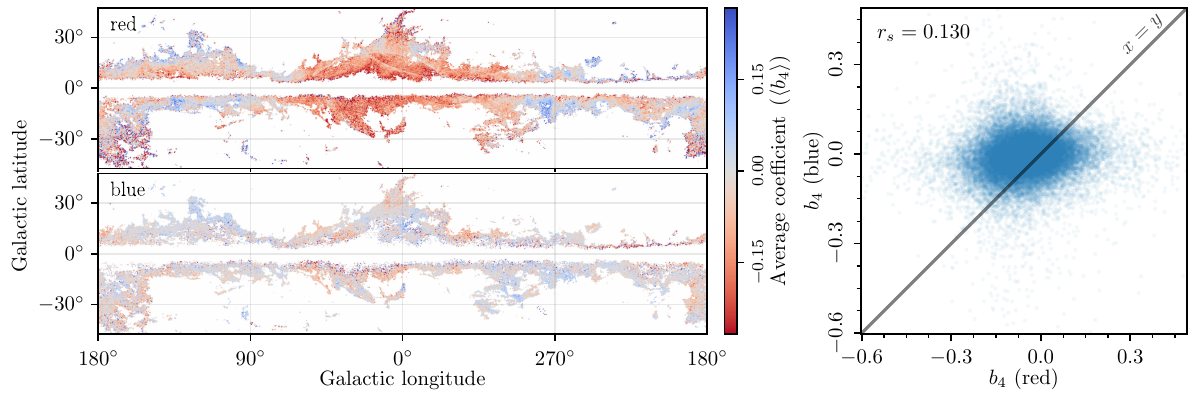}
  \caption{Tests of the robustness of each component, based on sky maps of the average coefficients using blue or red sources. Each row shows a different coefficient. In each row, the left panels show maps of the average coefficients, calculated using only red or blue sources, while the right panel shows the Spearman rank coefficient ($r_s$) between the individual pixels, with values from the ``red'' map on the $x$-axis, and values from the ``blue'' map on the $y$-axis. For illustrative purposes, we show only coefficients 0, 3 and 4. For components 0 and 3, the maps constructed using blue and red sources are highly correlated with one another and look visually similar. This is not the case with component 4, indicating that it is affected by color-dependent Gaia systematics. \label{fig:bluered_split}}
\end{figure*}

We will now further investigate whether our recovered extinction curve components correspond to real variations, or whether they result from systematics in the data. A number of studies have identified systematic errors in the Gaia XP spectra that depend on observed source color and apparent magnitude. \citet{Montegriffo2023GaiaDR3BPRPExternalCalibration} and \citet{Huang2024GDR3CorrectedXPSpectra} find color- and magnitude-dependent systematic errors that are particularly strong at the blue end of the spectrum ($\lambda < 400\,\mathrm{nm}$), but which also manifest as ``wiggles'' at the red end of the spectrum, increased residuals at $\lambda \approx 580\,\mathrm{nm}$, and a sharp break at the interface between the BP and RP spectra ($\lambda \approx 640\,\mathrm{nm}$). Because we do not use wavelengths shorter than 392\,nm, we should avoid the worst of these color-dependent systematics at the blue end of the spectrum.

The wavelength-dependence of the components can help identify the presence or absence of Gaia systematics. In the lowest-order components (0 -- 3), we do not see strong wiggle features at the blue or red end of the spectrum (left panels of Fig.~\ref{fig:skymap_components_0}). However, strong wiggles are present in many of the higher-order components (Figs.~\ref{fig:skymap_components_1}-\ref{fig:skymap_components_3}). At the interface of the BP and RP spectra ($\approx 650\,\mathrm{nm}$), we see sharp breaks in components 8 and 9. The sky map of component 8, in particular, also shows strong Gaia scanning-pattern artifacts. Together, these two properties suggest that component 8 (and possibly component 9) is contaminated by systematic differences in the relative flux calibration of BP and RP.

Extinction-curve components that arise from real variations in the dust properties should depend only on the dust itself, and not on the properties -- whether intrinsic (such as effective temperature) or observational (such as apparent magnitude) -- of the stars one observes. We can therefore test the robustness of each extinction-curve component to Gaia systematics by comparing sky maps generated using blue vs. red stars (based on observed color), faint vs. bright stars (based on apparent magnitude), hot vs. cool stars (based on inferred stellar temperature), and other splits on observed and intrinsic stellar parameters. If the sky map of a particular extinction-curve component depends strongly on observed color, for example, then that particular component is likely caused by systematics in the Gaia XP spectra, rather than real dust variations.

When constructing these pairs of maps, we must be careful to ensure that each map probes the same dust column. For example, a map based only on stars with red observed colors will tend to probe higher-extinction regions of the Galaxy than a map constructed using only stars with blue observed colors. We therefore ensure that all stars used to construct these map pairs probe nearly the \textit{entire} Milky Way dust column, by selecting stars that are more than 400\,pc from the Galactic midplane (at $> \! 2\sigma$ confidence):
\begin{align}
  \frac{\varpi_{\mathrm{est}}+2\sigma_{\varpi}}{1\,\mathrm{mas}}
  < \frac{\left|\sin b\right|}{0.4}
  \, ,
\end{align}
where $b$ is Galactic latitude, and $\varpi_{\mathrm{est}}$ and $\sigma_{\varpi}$ are the ZG25 estimates of stellar parallax and their uncertainties (based on Gaia parallax, XP spectra and NIR photometry from 2MASS and WISE). We use stars passing this cut to construct four pairs of maps, each splitting on a different Gaia observable or inferred stellar property:
\begin{itemize}
  \item \textbf{blue/red}: split at Gaia $BP-RP = 1.359\,\mathrm{mag}$ (the median color of stars passing our cuts).
  \item \textbf{faint/bright}: split at Gaia $m_G = 14.67\,\mathrm{mag}$ (the median apparent magnitude).
  \item \textbf{hot/cool}: split at $T_{\mathrm{eff}} = 6000\,\mathrm{K}$ (as inferred by ZG25).
  \item \textbf{metal-poor/rich}: split at $\left[\mathrm{Fe/H}\right] = -0.5\,\mathrm{dex}$ (as inferred by ZG25).
\end{itemize}
For each split, we generate pairs of inverse-variance weighted sky maps of the average strength of every component, using a HEALPix pixelization with $\mathtt{nside} = 128$ (corresponding to an angular resolution of $\sim 27^{\prime}$). We discard pixels with low extinction (inverse-variance-weighted $E < 0.15$) or fewer than 4 stars. We compare each pair of maps both qualitatively and quantitatively. Qualitatively, the spatial patterns in each pair of maps should match. In order to make a quantitative measurement, we calculate the Spearman rank coefficient ($r_s$) between the individual pixel values in the two maps. High correlation indicates that the information in the two maps is equivalent, and thus that the given component is not affected by the given split. Low correlation suggests that the given feature is due to systematics that manifest differently across the split (\textit{e.g.}, in blue vs. red sources).

We consistently find that the first four components (0 -- 3) pass both our qualitative and quantitative tests. That is, the strengths of these components are the same in blue vs. red sources, faint vs. bright sources, hot vs. cool stars, and metal-poor vs. metal-rich stars. However, all higher-order components are affected by at least one of the splits, indicating that they trace Gaia systematics to some extent. Fig.~\ref{fig:bluered_split} illustrates this using the blue/red split for components 0, 3 and 4. While the red/blue maps for components 0 and 3 match well (both qualitatively and quantitatively), the red/blue maps for component 4 look completely different from one another and have a low correlation. Appendix~\ref{app:split_tests} shows results for all of the splits, applied to all of the components.

One should expect that the correlation between each pair of maps should slowly fall off for increasingly high-order components, because these components represent weaker extinction-curve variations and therefore have noisier measurements. Indeed, we do see a slow decline in the correlations from component 0 to component 3. However, there is a sharp drop-off in correlation at component 4, suggesting that components 4 and above are largely unphysical. Visual inspection of the pairs of component maps is also informative, as it can differentiate between the case where two maps show the same spatial patterns but different statistical noise, and the case in which two maps show different spatial patterns. In the bottom-left panel of Fig.~\ref{fig:bluered_split}, for example, the red sky map of component 4 shows prominent Gaia scanning patterns that are absent in the blue sky map.

Finally, we note that the split tests we conduct here are insensitive to the choice of map resolution. Here, we use a HEALPix $\mathtt{nside} = 128$ pixelization. However, split tests using coarser map resolutions yield consistent results: components 0 -- 3 are robust against splits on color, \textit{etc}., while the higher-order components are not robust.

Based on the ISM-like patterns in the sky maps of components 0 -- 3, and the robustness of these maps against splits on source color and magnitude, and stellar temperature and metallicity, we conclude that these components capture real variation in the dust. There are thus at least four degrees of freedom that control the shape of the optical extinction curve, and these degrees of freedom are detectable in Gaia XP spectra.

\section{Distinct extinction features}
\label{sec:distinct-features}

\begin{figure}
  \centering
  \includegraphics[width=0.45\textwidth]{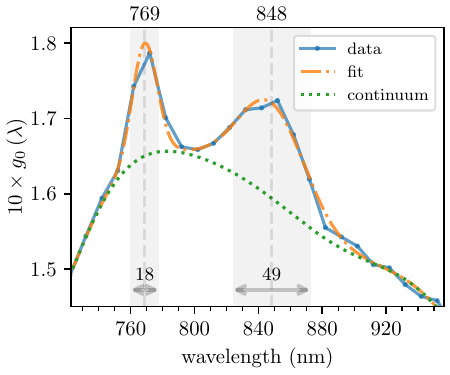}
  \caption{A joint fit of the parameters of the 770 and 850\,nm features. As these two features show up prominently in basis vector $0$, we model their properties in this basis vector, using the wavelength range 692--982\,nm. We treat each feature as a Gaussian profile, and model the continuum as a heavily regularized $11^{\mathrm{th}}$-order Chebyshev-polynomial expansion. Above, we plot the basis vector $\vec{g}_0$ (the data) and our fit. \label{fig:feature_fits}}
\end{figure}

\begin{figure}
  \centering
  \includegraphics[width=0.45\textwidth]{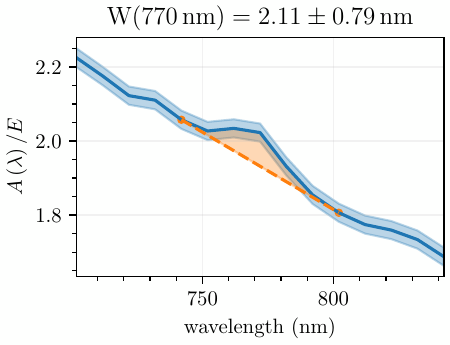}
  \caption{An illustration of the feature-width measurement process for an individual extinction curve. We draw a linear baseline between the edges of the extinction feature in question, as shown by the dotted orange line. The feature width is defined as the integral between the baseline and the normalized extinction curve (represented by the blue curve, with $1\sigma$ uncertainties shown in light blue). As the normalized extinction curve is unitless, the resulting integral has units of length. We measure the strength of three features in this manner: the ``very broad structure'' (VBS), and features at 770 and 850\,nm. \label{fig:linewidth_measurement}}
\end{figure}

\begin{figure*}
  \centering
  \includegraphics[width=0.95\textwidth]{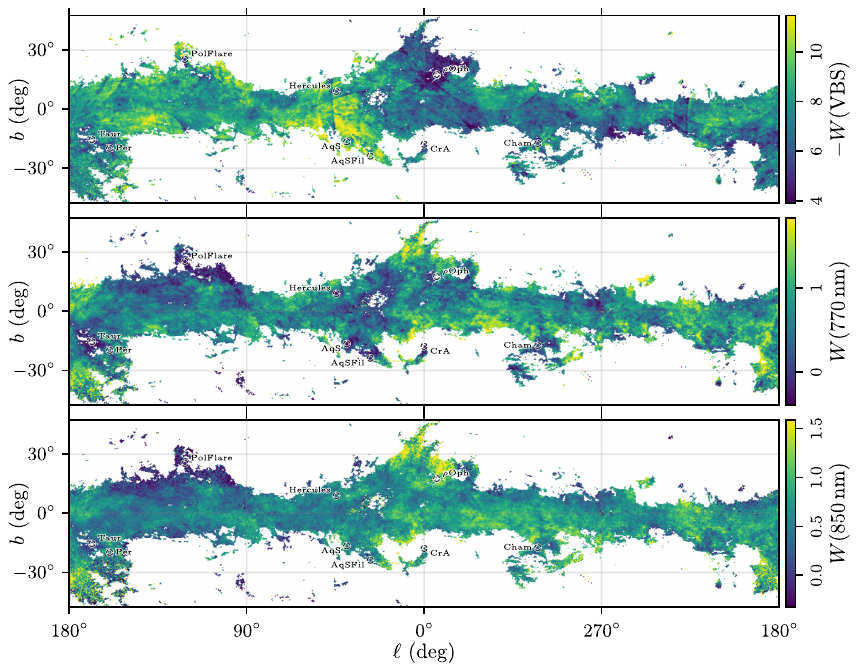}
  \caption{The inverse-variance-weighted mean equivalent width of three prominent features in our extinction curve: the ``very broad structure'' (VBS), which is a broad depression in extinction, and two bumps in extinction, at 770 and 850\,nm. The 770\,nm feature was previously identified by \citet{MaizApellaniz2021ISMBand7700AA} and detected in Gaia XP spectra by \citet{Zhang2024EmpiricalExtinctionCurve}, but the 850\,nm feature has not been previously identified. As the VBS is a depression in the extinction curve, we plot its negative equivalent width. All three features show rich structure over the sky. To orient the reader, we indicate the positions of several dust clouds: Taurus (Taur), Perseus (Per), the Polaris Flare (PolFlare), Hercules, Aquila South (AqS), filamentary structure near Aquila South (AqSFil), Corona Australis (CrA), $\rho$~Ophiuchus ($\rho$~Oph), and Chamaeleon (Cham). Note that while the maps of $W\!\left(770\,\mathrm{nm}\right)$ and $W\!\left(850\,\mathrm{nm}\right)$ are relatively free of Gaia scanning patterns, the map of $-W\!\left(\mathrm{VBS}\right)$ is not, possibly due to the greater level of noise in the Gaia XP spectra in the wavelength range of the VBS (see the left panels of Figs.~\ref{fig:skymap_components_1}--\ref{fig:skymap_components_3}). \label{fig:W_VBS_770_850}}
\end{figure*}

\begin{figure}
  \centering
  \includegraphics[width=0.45\textwidth]{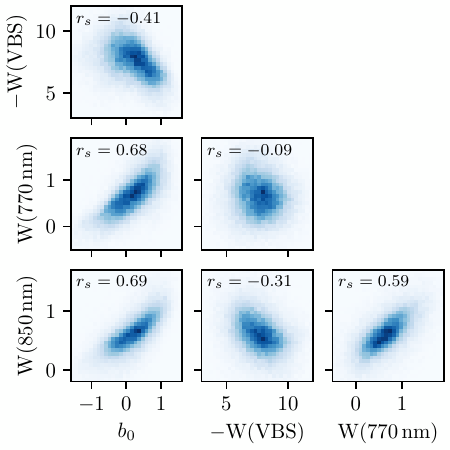}
  \caption{Correlations between the strengths of different features in our sky maps. We compare the amplitude $b_0$ of basis vector 0 (which represents $R(V)$ variation), and the equivalent widths of the VBS and the 770 and 850\,nm features. In each panel, we show the Spearman rank coefficient, $r_s$, between the two features. A value of $+1$ indicates perfect correlation, while $0$ indicates no correlation and $-1$ indicates perfect anti-correlation. The 770 and 850\,nm features are positively correlated with one another, and both are positively correlated with $R(V)$ (represented above by $b_0$). The strength of the VBS is anti-correlated with $R(V)$, and is slightly anti-correlated with the 850\,nm feature. \label{fig:feature_correlations}}
\end{figure}

\begin{figure*}
  \centering
  \includegraphics[width=0.95\textwidth]{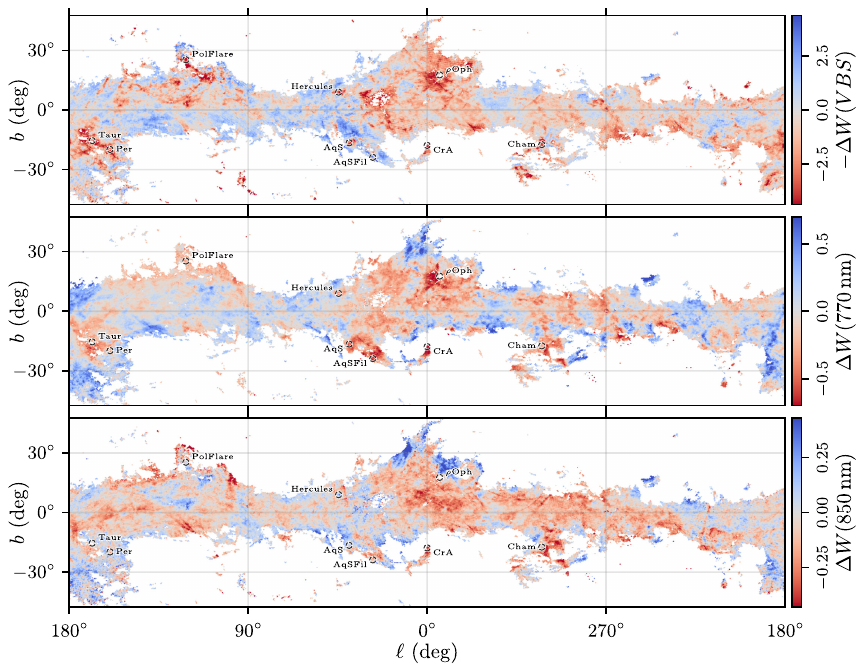}
  \caption{Sky maps of the mean difference in equivalent width of the VBS, 770\,nm and 850\,nm features from what one would expect, based on their correlation with $R(V)$. The strengths of these three features are not perfectly correlated with $R(V)$, and indeed we see spatial patterns in the residuals that have ISM-like structure. Notably, we see strong reductions in the equivalent widths of the VBS and 770\,nm features (but not of the 850\,nm feature) in the vicinity of $\rho$~Ophiuchus, a star-forming region with several O/B-star associations and a strong UV radiation field. This may hint at photodissociation of the carriers of these features. \label{fig:dW_VBS_770_850}}
\end{figure*}

\begin{figure*}
  \centering
  \includegraphics[width=0.95\textwidth]{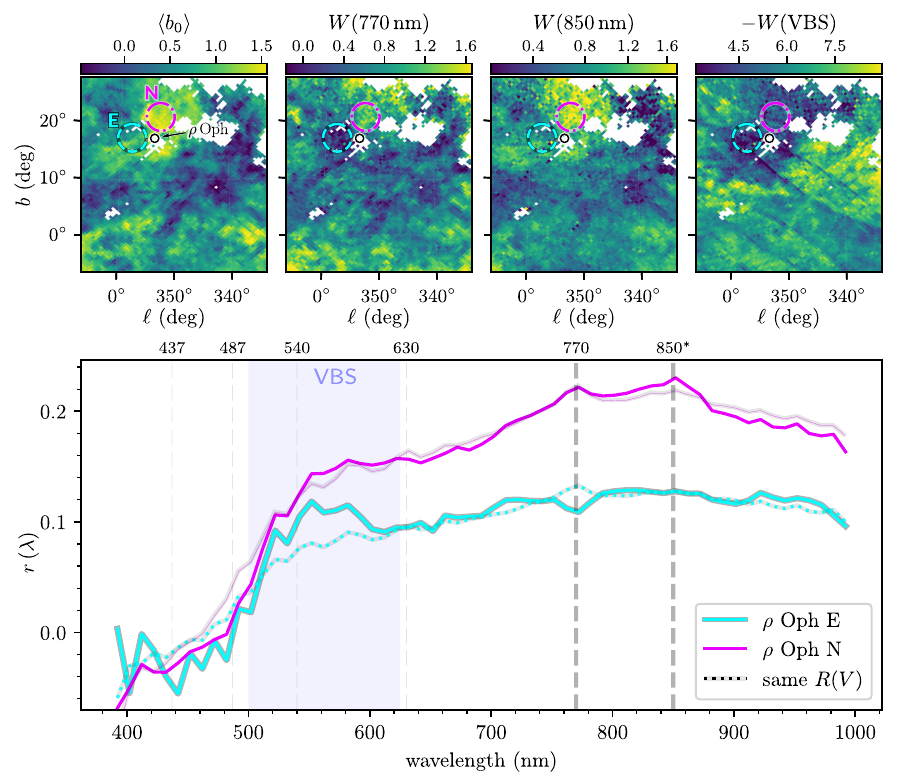}
  \caption{Top panels: Sky maps of $\langle b_0 \rangle$ (which traces $R(V)$), the 770 and 850\,nm features and the VBS in the direction of the inner Galaxy. The first three features trace similar -- but not identical -- structure, which has typical ISM-like filaments and cloud structures. The depth of the VBS is anti-correlated with the first three features, and contains imprints of Gaia scanning patterns. The largest differences between the maps are in the vicinity of $\rho$~Ophiuchus, where intense UV radiation from nearby O/B associations should strongly affect the dust chemistry. Bottom panel: The mean extinction curve (centered by subtracting the mean of the extinction curve across the entire sky) in two circled regions, North (N) and East (E) of $\rho$~Ophiuchus. The extinction curve in the region North of $\rho$~Ophiuchus shows a stronger 850\,nm feature than other areas of the sky with an equivalent $R(V)$ (shown by the dotted line of the same color). The region East of $\rho$~Ophiuchus shows a reduced 770\,nm feature compared to other regions of the sky with the same $R(V)$, as well as a shallower VBS (as can be seen from the difference between the solid and dotted cyan lines in the wavelength range of the VBS). The particular behavior in both of these regions illustrates that $R(V)$ is insufficient to fully describe the extinction curve, and suggests that the dust in different regions with the same $R(V)$ can have different chemical composition. \label{fig:skymaps_features_zoomin}}
\end{figure*}

Three distinct features show up strongly in our extinction curve components 0--3 (see the left panels of Fig.~\ref{fig:skymap_components_0}): the ``very broad structure'' (VBS) from roughly 500--625\,nm, and two additional extinction features, roughly centered on 770 and 850\,nm. The 770\,nm feature, discovered by \citet{MaizApellaniz2021ISMBand7700AA} and confirmed by \citet{Zhang2024EmpiricalExtinctionCurve}, shows up in all four of our most significant components, always accompanied by a second feature at 850\,nm, which has not been previously identified. As shown in Fig.~\ref{fig:feature_fits}, for the ``770\,nm'' feature, we measure a central wavelength of 769\,nm and a full-width at half-maximum (FWHM) of 18\,nm, both of which are consistent with the results found by \citet{MaizApellaniz2021ISMBand7700AA}. For the ``850\,nm'' feature, we measure a central wavelength of 848\,nm and a FWHM of 49\,nm -- more than twice as broad as the 770\,nm feature, which is itself far broader than any known DIB. For details of this fit, see Appendix~\ref{app:joint_770_850nm_feature_fit}.

We measure the strength of the VBS and the 770 and 850\,nm features in each normalized MQ extinction curve. As illustrated in Fig.~\ref{fig:linewidth_measurement}, we define a linear baseline below each feature (anchored at 442 and 802\,nm for the 770\,nm feature, and at 812 and 892\,nm for the 850\,nm feature), and then integrate the area between the baseline and the normalized extinction curve. The result has units of length, and is approximately equal to the equivalent width of the feature at an extinction of $E = 1$ (using the ZG25 extinction parameter). As this integral is a linear operation on the extinction curve, the uncertainty on feature width can be easily calculated from the covariance matrix of the extinction curve.

Fig.~\ref{fig:W_VBS_770_850} shows sky maps of the inverse-variance-weighted mean width of each feature on the sky. The correlations between these features are also of great interest, as they might hint that similar chemistry drives the features. Fig.~\ref{fig:feature_correlations} shows the correlations between our sky maps of four features: the amplitude $b_0$ of basis vector 0 (which represents $R(V)$ variation), and the equivalent widths of the VBS and the 770 and 850\,nm features. We see strong positive correlations between $R(V)$ (as measured by $b_0$) and the strength of the features at 770 and 850\,nm, and a strong positive correlation between the strengths of the 770 and 850\,nm features. We thus confirm the finding of \citet{MaizApellaniz2021ISMBand7700AA} that $R(V)$ and the equivalent width of the 770\,nm feature are correlated. In our maps, the depth of the VBS is moderately anti-correlated with $R(V)$, slightly anti-correlated with the strength of the 850\,nm feature, and has no detectable correlation with the 770\,nm feature.

Larger $R(V)$ corresponds to a flatter extinction curve, indicating a larger proportion of large vs. small dust grains. The positive correlation between $R(V)$ and the features at 770 and 850\,nm suggests that as small grains become relatively less abundant, the carriers of the 770 and 850\,nm features increase in abundance. In order to gain additional insight into the behavior of these features, we calculate the deviation of equivalent width from what one would expect, based on the correlation with $R(V)$. The 16-dimensional subspace representation of the normalized extinction curves is particularly convenient for this purpose. As the equivalent width is a linear transformation of the normalized extinction curve, it can also be expressed as a linear transformation of the coefficients $b_i$ of the basis vectors in our subspace. For example, including the contribution from the mean extinction curve ($R_{\mathrm{zp}}$), we find that the equivalent width of the 770\,nm feature is given by $W\!\left(770\,\mathrm{nm}\right) = 0.6342 + 0.3929 \, b_0 - 3.1316 \, b_1 + 2.4151 \, b_2 - 5.7142 \, b_3$, plus contributions from higher-order components (which we omit here for brevity). As basis vector 0 represents $R(V)$ variation, we can obtain the deviation of equivalent width from the trend with $R(V)$ by only including terms at order $b_1$ and higher: $\Delta W\!\left(770\,\mathrm{nm}\right) = - 3.1316 \, b_1 + 2.4151 \, b_2 - 5.7142 \, b_3 + \cdots$.

Fig.~\ref{fig:dW_VBS_770_850} plots $\Delta W$ for the VBS and 770 and 850\,nm features across the sky. The strengths of both the VBS and the 770\,nm feature are strongly reduced in the vicinity of $\rho$~Ophiuchus, while the 850\,nm features shows no large reduction in strength in this region. Fig.~\ref{fig:skymaps_features_zoomin} zooms in on the inner Galaxy, and shows the average extinction curve in two regions near $\rho$~Ophiuchus: one in which the 770\,nm feature is suppressed, and another in which the 850\,nm feature is stronger than other sightlines with comparable $R(V)$. One possible explanation for this behavior would be if the carrier of the 770\,nm feature undergoes photodissociation from the intense UV radiation environment in the $\rho$~Ophiuchus region. This finding is in line with \citet{MaizApellaniz2021ISMBand7700AA}, which found that the equivalent width of the 770\,nm feature is suppressed in star-forming regions with O/B associations. \citet{MassaFitzpatrickGordon2020OpticalExtinctionISS} argue that the VBS is the result of the extinction peaks at 487 and 630\,nm, representing a minimum between these two features. If this is true, a reduction in the depth of the VBS corresponds to a reduction in the strength of the 487 and 630\,nm features, and the carriers of these features may also be destroyed in regions with strong UV radiation.

We now pose the question of the reality of the 850\,nm feature. The 850\,nm feature shows up not only in Gaia XP extinction curves -- it can be seen (as can the 770\,nm feature) in Fig.~5 of \citet[][``G23'']{Gordon2023FUVtoMIRExtinctionCurve}, which plots the residuals between the measured and modeled mean extinction curve (which they label ``$a - \mathrm{fit}$''). G23 notes, however, that this region of the spectrum could be affected by the Paschen jump, which occurs at 820.4\,nm. There are three reasons why we do not believe the Paschen series to be responsible for the 850\,nm feature. The first reason is that the sky map of this feature shows ISM-like spatial patterns, and is indeed highly correlated with the sky map of the 770\,nm feature, which cannot be affected by the Paschen series, as well as the sky map of $b_0$, which traces $R(V)$ and should be fairly insensitive to the Paschen lines. The top panels of Fig.~\ref{fig:skymaps_features_zoomin} zoom in on these three features (and the VBS) near the Galactic Center, showing that they trace similar -- but not identical -- ISM-like structure. The bottom panel of Fig.~\ref{fig:skymaps_features_zoomin} shows the mean extinction curves in two regions in the vicinity of $\rho$~Ophiuchus where the relative strengths of the 770 and 850\,nm features differ, illustrating that these features are not perfectly correlated. The second reason is that restricting our stellar sample to low-temperature stars ($T_{\mathrm{eff}} < 6000\,\mathrm{K}$), for which the Paschen lines should be very weak, leaves the sky map of $W\left(850\,\mathrm{nm}\right)$ nearly unchanged. A third, related reason is that on a star-by-star basis, $W\left(850\,\mathrm{nm}\right)$ is uncorrelated with $T_{\mathrm{eff}}$. Using MQ extinction curves with uncertainty of less than 0.75\,nm in effective width, we measure a Spearman rank coefficient of -0.129 with $T_{\mathrm{eff}}$, as compared to much stronger coefficients of 0.592 with $R(V)$ and 0.342 with $W\left(770\,\mathrm{nm}\right)$ (note that this star-by-star comparison is noisier than the comparisons shown in Fig.~\ref{fig:feature_correlations}, which are based on sky maps of average feature strengths). Finally, we note that the 850\,nm feature can also be seen in Fig.~5 of \citet[][``Zh24'']{Zhang2024EmpiricalExtinctionCurve}, and is particularly clear in the residuals between their model and that of G23. While the Zh24 extinction curves are also measured using XP, and therefore share the same Gaia systematics as our extinction curves, their method of determining the intrinsic stellar spectrum (\textit{i.e.}, pairing with unreddened analogs using LAMOST spectra) is completely different.

\section[Discussion of the origins of the 770 and 850 nm features]{Discussion of the origins of the 770 and 850\,nm features}
\label{sec:origins_770_850}

\citet[][``ZHG25'']{ZhangHensleyGreen2025RvPAHs} argues that much of the $R(V)$ variation in the translucent ISM of the Milky Way and LMC is driven by variations in PAH abundance. PAHs are thought to be abundant in the ISM \citep[\textit{e.g.},][]{Wenzel2024OneCyanopyreneTMC1,Wenzel2024TwoFourCyanopyreneTMC1}, and generically contain an electronic $\pi \leftrightarrow \pi^{*}$ transition in the UV that may be the origin of the 2175\,\AA{} feature \citep{Draine2003InterstellarDust}. Though the center of this feature is in the UV, its long-wavelength wing extends into the optical, where it steepens the extinction curve. An increase in PAH mass fraction (compared to the total dust mass) from 4\% to 8\%, which is allowed by the carbon budget in the gas phase of the diffuse ISM, is sufficient to reduce $R(V)$ from 3.4 to 2.9 \citep{ZhangHensleyGreen2025RvPAHs}, and indeed just such a trend is seen in several clouds studied by ZHG25.

In this picture, destruction of PAHs (for example, by UV radiation) increases $R(V)$, while growth of PAHs through accretion of carbon from the gas phase causes $R(V)$ to decrease \citep{ZhangHensleyGreen2025RvPAHs}. In the present work, we see an increase in the strength of the 770 and 850\,nm features as $R(V)$ increases (see Fig.~\ref{fig:feature_correlations}). This lends itself to the conjecture that the carriers of the 770 and 850\,nm features are small carbon-bearing molecules that become more abundant in environments where PAHs are destroyed. As seen in Figs.~\ref{fig:dW_VBS_770_850} and \ref{fig:skymaps_features_zoomin}, the 770\,nm feature is suppressed in the vicinity of $\rho$~Ophiuchus, which is a region with several O/B associations and thus a particularly strong UV radiation field. In this environment, the carrier (or carriers) of the 770\,nm feature may also undergo photodissociation. The 850\,nm feature shows no such suppression near $\rho$~Ophiuchus, possibly indicating that its carrier (or class of carriers) is more stable.

\citet{MaizApellaniz2021ISMBand7700AA} found that the 770\,nm feature is weaker along sightlines that are rich in diatomic carbon, $C_2$, indicating some connection to carbon chemistry. However, absent specifically identified carriers with transitions at the correct energies, and absent identification of additional transitions belonging to these carriers (\textit{e.g.}, rotational transitions in the radio spectrum), this conjecture is highly tentative.

\section{Conclusions}
\label{sec:conclusions}

Though the optical extinction curve has commonly been assumed to be described by a single parameter, $R(V)$, the availability of large numbers measured extinction curves -- determined using low-resolution Gaia XP spectra -- reveals additional physical degrees of freedom that are richly structured on the sky.

In this paper, we identify a new, broad optical extinction feature centered on $\lambda \approx 848\,\mathrm{nm}$ (the ``850\,nm'' feature), with a FWHM of 49\,nm (several times the width of the broadest known DIB). We measure the equivalent widths of the VBS and the 770 and 850\,nm features in 24 million extinction curves, allowing us to map the strength of each feature across the sky. These features show strong correlations with one another and with $R(V)$. However, $R(V)$ is insufficient to predict the strength of these features in every environment. For example, in the vicinity of $\rho$~Ophiuchus, where there are a number of O/B associations and a history of recent ($\sim$20\,Myr) star formation \citep{Ratzenboeck2023ScoCenStarFormation}, the 770\,nm feature is much weaker than one would expect, based on its overall correlation with $R(V)$, indicating possible photodissociation of its carrier by intense UV radiation. By contrast, the 850\,nm feature does not show a deficit in this same region.

Our extinction curve decompositions and equivalent-width measurements for the VBS and 770 and 850\,nm features are available at \href{https://dx.doi.org/10.5281/zenodo.14005028}{DOI:10.5281/zenodo.14005028}. The large catalog, densely covering the entire Galactic plane, will allow more detailed investigation of connections between optical extinction features and ISM conditions and chemistry.

\section*{Acknowledgments}

GG and XZ are supported by a Sofja Kovalevskaja Award to GG from the Alexander von Humboldt Foundation. We have had useful conversations about some of the results in this paper with Brandon Hensley, Christiaan Boersma, Vincent Esposito and Alexandros Maragkoudakis.

This work presents results from the European Space Agency (ESA) space mission Gaia. Gaia data are being processed by the Gaia Data Processing and Analysis Consortium (DPAC). Funding for the DPAC is provided by national institutions, in particular the institutions participating in the Gaia MultiLateral Agreement (MLA). The Gaia mission website is https://www.cosmos.esa.int/gaia. The Gaia archive website is https://archives.esac.esa.int/gaia.


\bibliography{bibliography}{}
\bibliographystyle{aasjournal}



\appendix

\section{Effect of observational uncertainties on estimated extinction}
\label{app:lnx-derivation}

When both the intrinsic and observed ($f_{\mathrm{obs}}$) flux are known exactly, extinction is simply given by
\begin{align}
  A = -2.5 \log_{10} \left(
    \frac{f_{\mathrm{obs}}}{f_{\mathrm{int}}}
  \right) \, .
\end{align}
However, when $f_{\mathrm{obs}}$ contains observational uncertainties, there are corrections to the above formula. When the observational uncertainties are Gaussian, these corrections take the form of a power series in the inverse of the signal-to-noise ratio, $\sigma_f/\mu_f$. To see why this is, consider the expected value of $y = \ln\left(x\right)$, where $x$ is a Gaussian random variable with mean $\mu$ and standard deviation $\sigma$. Defining $z \equiv \left(x-\mu\right)/\sigma$,
\begin{align}
  \langle y \rangle
  = \langle \ln\left(x\right) \rangle
  = \langle \ln\left(\mu + \sigma z\right) \rangle
  = \ln\mu + \langle \ln\left(1 + \frac{\sigma}{\mu} z\right) \rangle
  \, .
\end{align}
Expanding the natural logarithm as a Taylor series,
\begin{align}
  \langle y \rangle
  = \ln\mu - \sum_{k=1}^{\infty}
    \frac{\left(-1\right)^k}{k}
    \left(\frac{\sigma}{\mu}\right)^{\! k}
    \langle z^k \rangle
  \, .
\end{align}
As $z$ is a unit Gaussian random variable, its odd moments are zero, and its even moments are given by $\left(k-1\right)!!$. Thus,
\begin{align}
  \langle y \rangle
  &= \ln\mu \ - \!\!\!\! \sum_{k=2,4,6,\ldots} \!\!\!\!
    \frac{\left(-1\right)^k}{k}
    \left(\frac{\sigma}{\mu}\right)^{\! k}
    \left(k-1\right)!!
  \\
  &= \ln\mu - \sum_{n=1}^{\infty}
    \frac{\left(2n-1\right)!!}{2n}
    \left(\frac{\sigma}{\mu}\right)^{\! 2n}
  \, .
  \label{eqn:lnx_series_result}
\end{align}
Applying this formula to the definition of extinction and keeping only the lowest-order term in the series expansion yields Eq.~\ref{eqn:A_mean}. The series in Eq.~\ref{eqn:lnx_series_result} is asymptotic, and diverges as $n\rightarrow\infty$. However, for reasonably small $\sigma/\mu$, it becomes more accurate as one adds in low-order terms. In our dataset, the per-wavelength flux signal-to-noise ratio (``SNR'', defined as $\mu/\sigma$) is generally greater than 10, with decreasing SNR at the blue end of the spectrum. At 400\,nm, only 1\% of the spectra in our MQ dataset have $\mathrm{SNR} < 10$, while at our shortest wavelength, 392\,nm, 1\% of spectra have $\mathrm{SNR} < 4$. For reference, with $\mathrm{SNR} = 5$, the first-correction to the extinction would be $\approx 0.02\,\mathrm{mag}$, which is small but non-negligible. The second-order correction is suppressed by an additional factor of SNR, and is thus negligible in our dataset.

A similar approach can be used to obtain the covariance $\mathrm{cov}\!\left( y_1, y_2 \right)$, where $y_i = \ln\left(x_i\right)$, and $x_1$ and $x_2$ are correlated Gaussian random variables:
\begin{align}
  \mathrm{cov}\!\left( y_1, y_2 \right) \simeq \left(
    \frac{\sigma_{1,2}^2}{\mu_1 \mu_2}
    -\frac{1}{4}\frac{\sigma_1^2 \sigma_2^2}{\mu_1^2 \mu_2^2}
  \right) \, ,
\end{align}
where $\sigma_{1,2}^2$ is the covariance between $x_1$ and $x_2$.

An entirely different approach would be to forward-model the spectral flux, which has a Gaussian likelihood. However, this would be incompatible with the fast CovPCA linear-algebra methods that we use here. We therefore treat the probability density of $A$ as Gaussian, and restrict our analysis to relatively high-SNR sources, for which this is a decent approximation.

\section{CovPCA Algorithm}
\label{app:cov_pca}

We have observations of $n_s$ extinction curves, $\vec{A}_i$ ($i$ indexes the curves). Each extinction curve is measured at $n_{\lambda}$ wavelengths. For each source, the covariances between the extinction errors at different wavelengths are described by the covariance matrix $\mathbf{C}_i$. We want to find a set of $n_b$ basis vectors ($n_b \ll n_{\lambda}$), represented by the basis matrix $G_{jk}$, that minimizes $\chi^2$:
\begin{align}
\chi^2 = \sum_{i, j, k, l, m} (G_{jk} B_{i, k} - A_{i,j})(C_{i}^{-1})_{jl}(G_{lm} B_{i, m} - A_{i,l}),
\end{align}
where $\vec{B}_i$ is the coefficient representation of extinction curves in the space spanned by $\mathbf{G_{n_{\lambda}\times n_b}}$. However, it is difficult to simultaneously determine $\mathbf{G_{n_{\lambda}\times n_b}}$ and $\mathbf{B_{n_b\times n_s}}$. We instead adopt an iterative method, similar to \citet{TsalmantzaHogg2012PCASpectra}. 

\subsection{B-step: Assume matrix G is known.}

To minimize $\chi^2$ w.r.t. the coefficients, $\mathbf{B}$, holding the basis vectors fixed, we require
\begin{align}
  \frac{\partial \chi^2}{\partial B_{i, p}}=0 .
\end{align}
\begin{align}
  \implies
  0 = \sum_{j, k, l, m} \left[
    G_{jp}(C_{i}^{-1})_{jl}(G_{lm} B_{i, m} - A_{i,l})
  + (G_{jk} B_{i, k} - A_{i,j})(C_{i}^{-1})_{jl}G_{lp}
  \right] \, .
\end{align}

If we re-label the dummy indices of the second term: ($j \rightarrow l$, $l \rightarrow j$, $k \rightarrow m$), we find it is identical with the first term:
\begin{align}
  0 = \sum_{j, k, l, m} \left[
    G_{jp}(C_{i}^{-1})_{jl}(G_{lm} B_{i, m} - A_{i,l})
  + (G_{lm} B_{i, m} - A_{i,l})(C_{i}^{-1})_{jl}G_{jp}
  \right] \, ,
\end{align}
\begin{align}
  \sum_{j, k, l, m} G_{jp}(C_{i}^{-1})_{jl}(G_{lm} B_{i, m} - A_{i,l}) = 0 \, .
\end{align}
In matrix notation, this is:
\begin{align}
  \mathbf{G}^T \mathbf{C}_i^{-1} \mathbf{G} \vec{B}_i
  = \mathbf{G}^T \mathbf{C}_i^{-1} \vec{A}_i
\end{align}
Note that $\mathbf{G}^T \mathbf{C}_i^{-1} \mathbf{G}$ is always invertible. Thus, the optimal coefficients are given by
\begin{align}
  \vec{B}_i =
    (\mathbf{G}^T \mathbf{C}_i^{-1} \mathbf{G})^{-1}
    \mathbf{G}^T \mathbf{C}_i^{-1} \vec{A}_i
   \, .
\end{align}

\subsection{G-step: Assume matrix B is known.}

Similarly, in order to minimize $\chi^2$ w.r.t. the basis vectors, holding the coefficients constant, we require, 
\begin{align}
  \frac{\partial \chi^2}{\partial G_{pq}} &= 0 \, .
\end{align}
\begin{align}
  \implies
  0 &= \sum_{i, j, k, l, m} \left[
    B_{i, q} (C_{i}^{-1})_{pl}(G_{lm} B_{i, m} - A_{i,l})
  + (G_{jk} B_{i, k} - A_{i,j})(C_{i}^{-1})_{jp} B_{i, q}
  \right]
  \, .
\end{align}

The second term is also the same as the first term if we re-label the dummy indices ($j \rightarrow l$, $k \rightarrow m$). Thus,
\begin{align}
  0 = \sum_{i, l, m}  B_{i, q} (C_{i}^{-1})_{pl} (G_{lm} B_{i, m} - A_{i,l})\, .
\end{align}
This gives us $\mathbf{n_{\lambda}\times n_b}$ equations.

We flatten the G matrix into a vector: $\mathbf{G_a^\prime}$
\begin{align}
  G^\prime_a = G_{lm} ,
\end{align}
where $m = \mathrm{floor}\,(a/n_{\lambda})$, $l = a\ \mathrm{mod}\ n_{\lambda}$. 

We can also flatten other matrices by $q = \mathrm{floor}(b/n_{\lambda})$, $p = b\ \mathrm{mod}\ n_{\lambda}$:
\begin{align}
  A_{ba}^\prime G_{a}^\prime &= R_b^\prime \, , \\
  A^\prime_{ba} &= \sum_i B_{i, q} (C_{i}^{-1})_{pl} B_{im} \, , \\
  R^\prime_b &= \sum_{i, l} B_{i, q} (C_{i}^{-1})_{pl} A_{il} \, .
\end{align}

Therefore, we can calculate $G^\prime$ by
\begin{align}
  G^\prime = (A^\prime)^{-1}R^\prime \, .
\end{align}

Reshaping $G^{\prime}$ yields the basis matrix, $\mathbf{G}$.

\subsection{Covariance matrix of basis vectors}

We can calculate the covariance matrix of each basis vector, $\vec{B}$, using the fact that it is a linear transformation of the corresponding vector $\vec{A}$:
\begin{align}
  \vec{B}
  &= \underbrace{\big(
      \overbrace{
        \mathbf{G}^T\mathbf{C}^{-1}\mathbf{G}
      }^{\equiv \, \mathbf{J}}
    \big)^{-1}
    \mathbf{G}^T\mathbf{C}^{-1}
  }_{\equiv \, \mathbf{K}}
  \vec{A}
  = \mathbf{K} \vec{A}
  \, .
\end{align}
Note that $\mathbf{J} = \mathbf{J}^T$. The covariance matrix of $\vec{B}$ is then given by
\begin{align}
  C_B
  &= \mathbf{K}\mathbf{C}\mathbf{K}^T
  = \mathbf{J}^{-1}\mathbf{G}^T
    \underbrace{\mathbf{C}^{-1}\mathbf{C}}_{=\,\mathbf{1}}
    \left(\mathbf{J}^{-1}\mathbf{G}^T\mathbf{C}^{-1}\right)^T
  = \mathbf{J}^{-1}
    \underbrace{
      \mathbf{G}^T\mathbf{C}^{-1}\mathbf{G}
    }_{=\,\mathbf{J}}
    \mathbf{J}^{-1}
  = \mathbf{J}^{-1}
  \\
  &= \left(\mathbf{G}^T\mathbf{C}^{-1}\mathbf{G}\right)^{-1}
  \, .
\end{align}

\section[Comparison with Massa et al. (2020)]{Comparison with Massa \textit{et al.} (2020)}
\label{app:massa_etal_2020_comparison}

\begin{figure}
  \centering
  \includegraphics[width=0.55\textwidth]{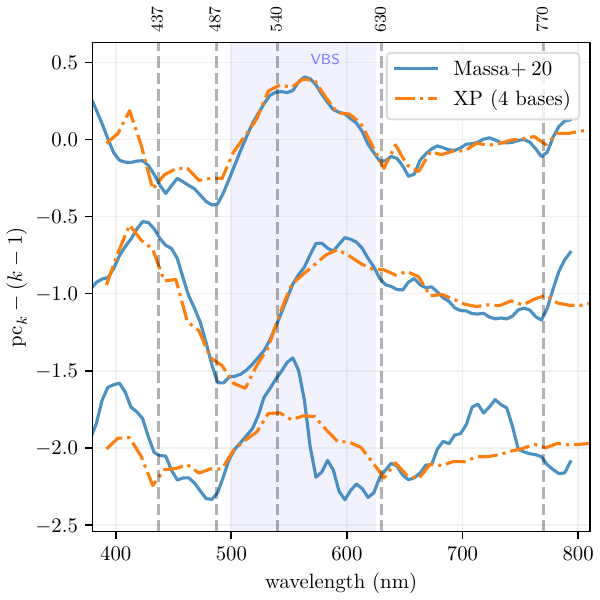}
  \caption{Comparison of the extinction curve principal components found by \citet[][``MFG20'']{MassaFitzpatrickGordon2020OpticalExtinctionISS} (in order of most to least significance from top to bottom) with our lowest-order basis vectors. For each MFG20 principal component, we overplot the best-matching linear combination of our four most significant basis vectors (labeled ``XP''). Our four most significant basis vectors capture the variation encoded by MFG20's $\mathrm{pc}_1$ and $\mathrm{pc}_2$, but not the variation encoded by $\mathrm{pc}_3$. This latter variation is distributed over our higher-order basis vectors. Note that the wavelength range of our basis vectors and that of MFG20's principal components are not identical -- we plot the overlap region. \label{fig:covpca_massa_comparison}}
\end{figure}

\citet[][``MFG20'']{MassaFitzpatrickGordon2020OpticalExtinctionISS} applied principal component analysis to extinction curves calculated from observed Hubble/STIS optical spectra, obtaining three statistically significant components. We compare these principal components to our basis vectors, which are obtained by applying CovPCA to the Gaia XP extinction curves.

Because the principal components of MFG20 could be a linear combination of our basis vectors, we find the combination of our basis vectors that minimizes the mean square difference with each MFG20 principal component, $\mathrm{pc}_k$, $k = 1, 2, 3$. As our space of basis vectors is very large (16 dimensions) compared to the three principal components of MFG20, we only use our four most significant basis vectors in this comparison. This essentially tests whether the same extinction-curve variations seen by MFG20 are captured by our lowest-order basis vectors. We find the following transformations:
\begin{align}
  \mathrm{pc}_1 &\simeq +0.0900\,\vec{g}_0 +0.3448\,\vec{g}_1 -0.2283\,\vec{g}_2 -0.2400\,\vec{g}_3 \, , \\
  \mathrm{pc}_2 &\simeq -0.1514\,\vec{g}_0 +0.4230\,\vec{g}_1 +0.4646\,\vec{g}_2 -0.0828\,\vec{g}_3 \, , \\
  \mathrm{pc}_3 &\simeq -0.0152\,\vec{g}_0 +0.1550\,\vec{g}_1 -0.1868\,\vec{g}_2 -0.2090\,\vec{g}_3 \, . \\
  \label{eqn:mfg20_transformation}
\end{align}
Fig.~\ref{fig:covpca_massa_comparison} plots our transformed basis vectors on top of the MFG20 principal components. We recover the first two principal components of MFG20 fairly closely, but the third principal component of MFG20 evidently contains extinction-curve variation that cannot be captured by our lowest-order principal components. Interestingly, the 770\,nm feature is visible in the first two MFG20 principal components, although it was only identified later, by \citet{MaizApellaniz2021ISMBand7700AA}.

\section{Results of all split tests}
\label{app:split_tests}

\begin{figure}
  \centering
  \includegraphics[width=0.85\textwidth]{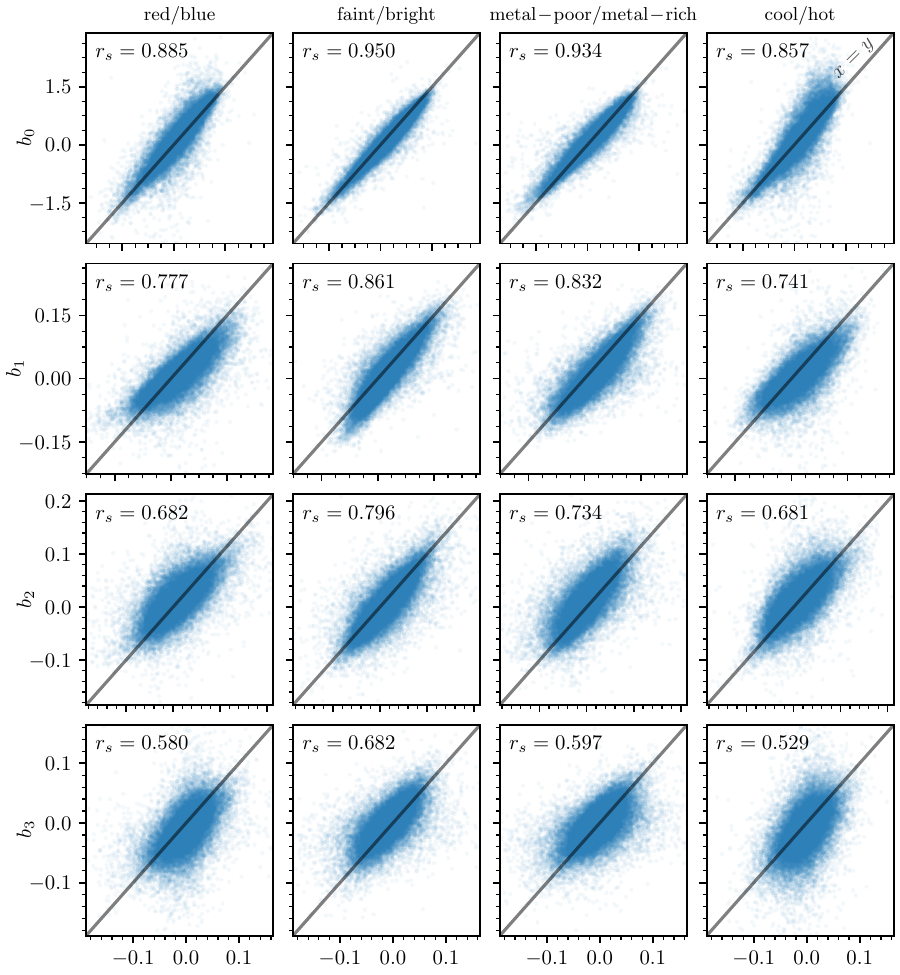}
  \caption{Correlations between sky maps of coefficients 0--3 (one per row) constructed using splits on color (1$^{\rm st}$ column, with the ``red'' map on the $x$-axis), apparent magnitude (2$^{\rm nd}$ column, ``faint'' map on the $x$-axis), metallicity (3$^{\rm rd}$ column, ``metal-poor'' map on the $x$-axis) and temperature (4$^{\rm th}$ column, ``cool'' map on the $x$-axis). In each panel, we show the Spearman rank coefficient ($r_s$) of the given pair of maps, with $r_s=1$ indicating perfect correlation, and $r_s=0$ indicating that the maps are completely uncorrelated. \label{fig:splits_scatter_0to3}}
\end{figure}

\begin{figure}
  \centering
  \includegraphics[width=0.85\textwidth]{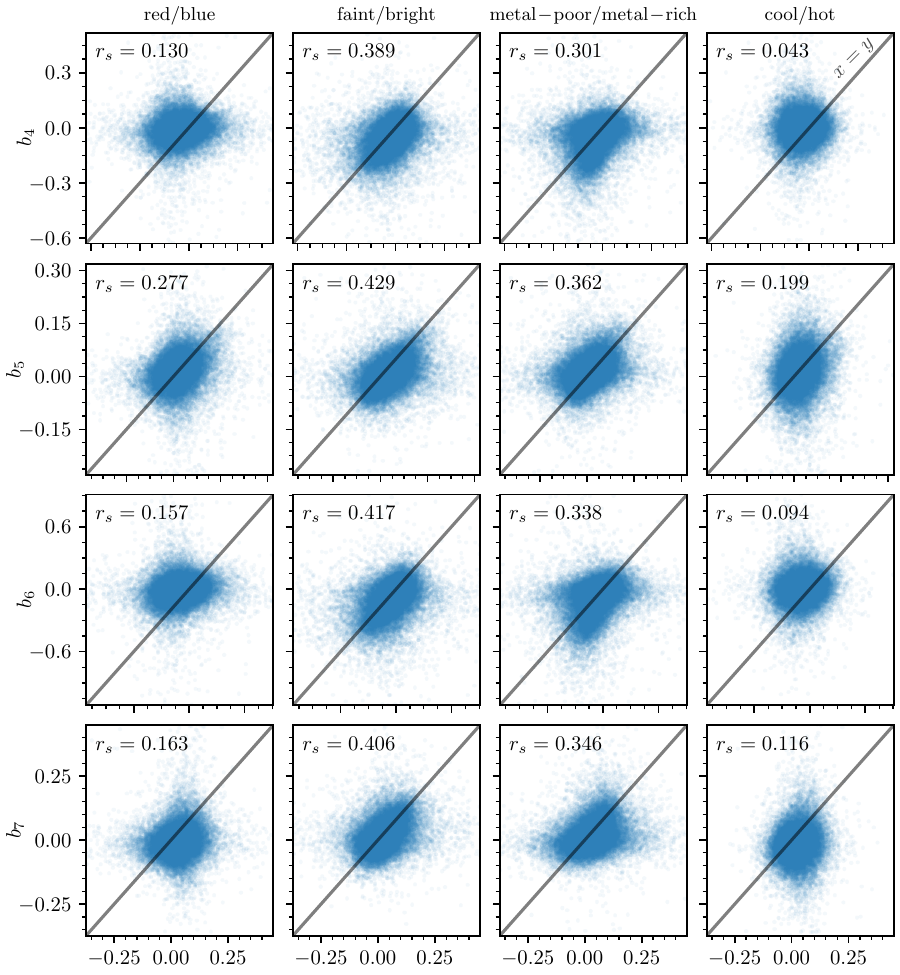}
  \caption{As Fig.~\ref{fig:splits_scatter_0to3}, but for coefficients 4--7. \label{fig:splits_scatter_4to7}}
\end{figure}

\begin{figure}
  \centering
  \includegraphics[width=0.85\textwidth]{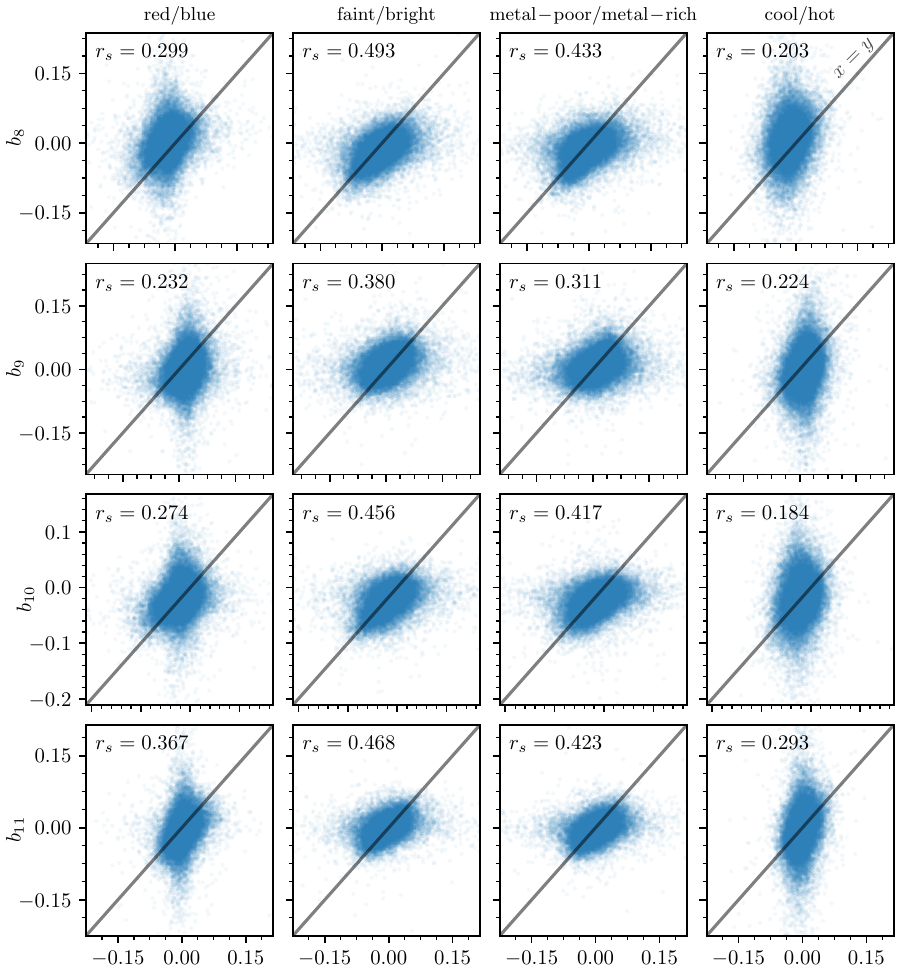}
  \caption{As Fig.~\ref{fig:splits_scatter_0to3}, but for coefficients 7--11. \label{fig:splits_scatter_8to11}}
\end{figure}

\begin{figure}
  \centering
  \includegraphics[width=0.85\textwidth]{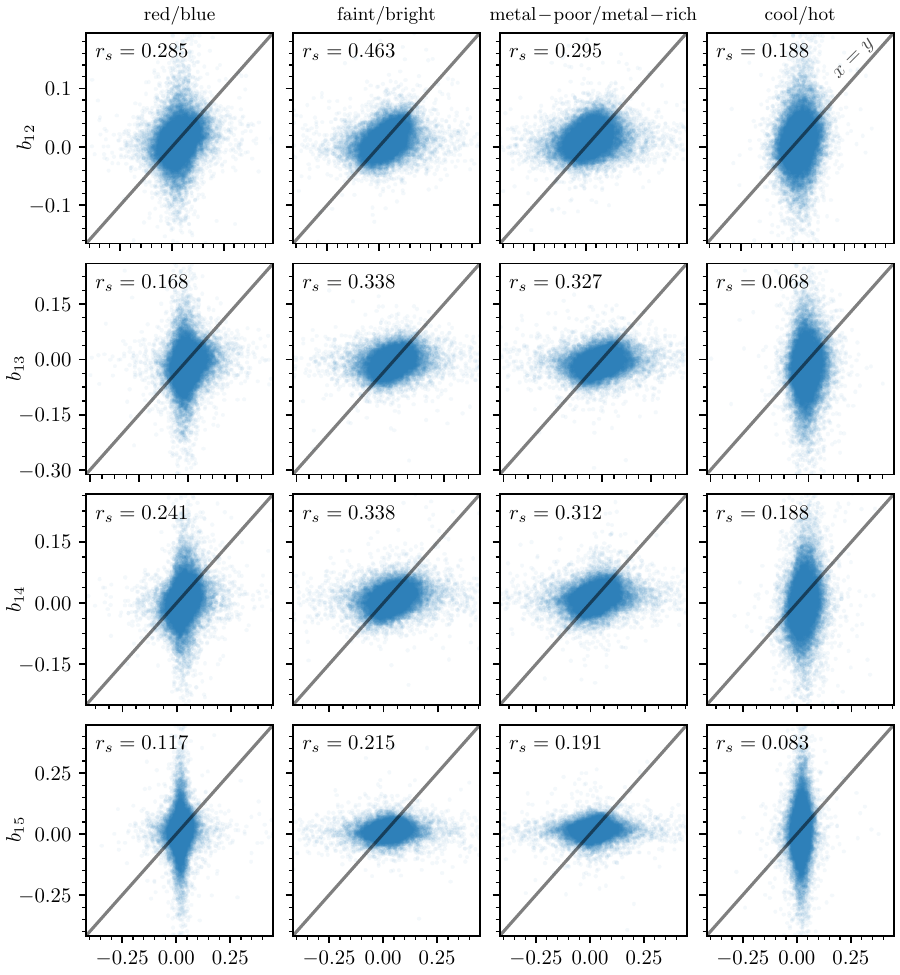}
  \caption{As Fig.~\ref{fig:splits_scatter_0to3}, but for coefficients 12--15. \label{fig:splits_scatter_12to15}}
\end{figure}

\begin{figure*}
  \centering
  \includegraphics[width=\textwidth]{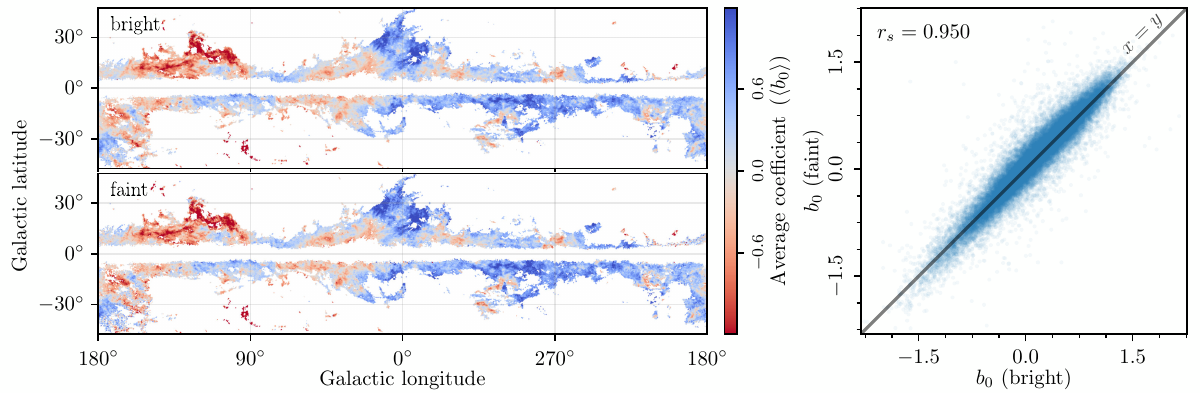}
  \includegraphics[width=\textwidth]{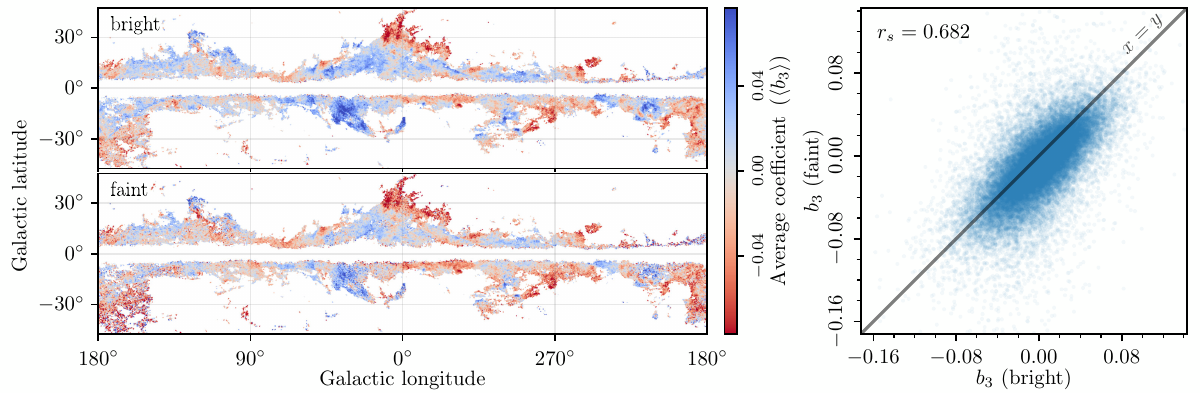}
  \includegraphics[width=\textwidth]{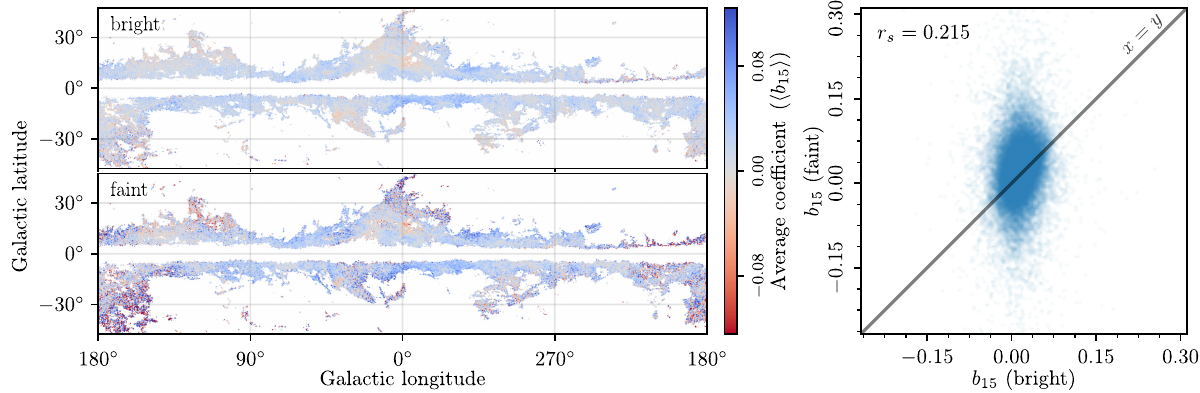}
  \caption{As with Fig.~\ref{fig:bluered_split}, but for faint vs. bright sources, and showing coefficient 15 instead of coefficient 4. For coefficients 0 and 3, the maps of constructed using faint and bright sources are highly correlated with one another and look visually similar. In contrast, coefficient 15 is likely affected by magnitude-dependent Gaia systematics, as can be seen from the low correlation between the faint and bright maps (bottom-right panel) and the checkerboard patterns in the faint map, which are absent from the bright map (bottom-left panel). \label{fig:faintbright_split}}
\end{figure*}

As explained in Section~\ref{sec:robustness}, we test the robustness of each component to Gaia systematics by comparing pairs of sky maps of average coefficient values, constructed using different splits on Gaia observables or stellar parameters. Figs.~\ref{fig:splits_scatter_0to3}, \ref{fig:splits_scatter_4to7}, \ref{fig:splits_scatter_8to11} and \ref{fig:splits_scatter_12to15} show the correlation of each pair of maps (for each combination of component and split). Fig.~\ref{fig:faintbright_split} shows sky maps of the faint/bright split for two components (0 and 3) that pass this test and one component (15) that performs particularly poorly on this test.

The first four components (0 -- 3) show high correlations ($r_s > 0.5$) across all splits, indicating that they are not strongly affected by Gaia systematics. There is a steep drop-off in correlation at component 4, and indeed all higher-order components have far lower correlations, indicating strong contamination by systematics. Of these higher-order components, component 11 is the most robust, with correlations of $r_s \geq 0.293$ across the different splits. As can be seen in the bottom-left panel of Fig.~\ref{fig:skymap_components_2}, component 11 alters the relative strength of the 770\,nm vs. the 850\,nm feature, but also contains ``wiggles'' that are likely unphysical.

\section[Details of joint 770 and 850 nm feature fit]{Details of joint 770 and 850\,nm feature fit}
\label{app:joint_770_850nm_feature_fit}

As illustrated in Fig.~\ref{fig:feature_fits}, we perform a joint fit of the properties of the 770 and 850\,nm features in our basis vector 0 (which represents $R(V)$ variation). This requires us to also model the continuum of basis vector 0. We model each feature as a Gaussian profile (with amplitude $a$, central wavelength $\bar{\lambda}$, and width $\Delta\lambda$), and the continuum as an 11$^{\mathrm{th}}$-order Chebyshev-polynomial expansion:
\begin{align}
  r_{\mathrm{pred}}\left(\lambda\right) \simeq
  \sum_{i=1}^2 a_i \exp\left[
    -\frac{
      \left(\lambda-\bar{\lambda}_i\right)^2
    }{
      2\left(\Delta\lambda_i\right)^2
    }
  \right]
  + \sum_{i=0}^{11} c_i \, T_i \left(
    \frac{
      \lambda-837\,\mathrm{nm}
    }{
      145\,\mathrm{nm}
    }
  \right)
  \, ,
\end{align}
where $T_i$ is the $i^{\mathrm{th}}$-order Chebyshev polynomial, $c_i$ is the corresponding fitting coefficient, and we model the range 792--982\,nm. We first obtain a rough fit by minimizing $\langle\left[r_{\mathrm{pred}}\left(\lambda\right)-g_0\left(\lambda\right)\right]^2\rangle$, holding $c_6$ through $c_{11}$ fixed at zero. This leads to a fit with a root-mean-square error of $\sigma_{\mathrm{rms}} = 5.56 \times 10^{-4}$.

Visual inspection shows that the resulting continuum fit contains moderately strong undulations that appear unphysical. We then re-run the fit, using the rough fit as a starting point, but allowing the higher-order Chebyshev polynomial coefficients to vary, and heavily regularizing all terms above order $c_2$. We minimize the objective function
\begin{align}
  \mathcal{L} =
  \frac{1}{2\sigma_{\mathrm{rms}}^2} \sum_{i=1}^{n_{\lambda}} \left[
    r_{\mathrm{pred}}\left(\lambda_i\right)
    - g_0\left(\lambda_i\right)
  \right]^2
  + 10^6 \sum_{i=3}^{11} c_i^2
  \, ,
\end{align}
where $n_{\lambda} = 30$ is the number of modeled wavelengths, and $\lambda_i = 692\,\mathrm{nm}$, $702\,\mathrm{nm}$, $\ldots$, $982\,\mathrm{nm}$. We obtain the fit shown in Fig.~\ref{fig:feature_fits}.

The central wavelengths and FWHM of the 770 and 850\,nm features are largely insensitive to the details of the continuum fit, and change by less than 1\,nm from our rough fit to our final fit. However, the amplitudes of the features are affected by the details of the continuum fit.

\end{document}